\documentclass[12pt,german,english]{article}
\usepackage[T1]{fontenc}
\usepackage[latin9]{inputenc}
\usepackage{geometry}
\usepackage{amsmath}
\usepackage{amssymb}
\usepackage{setspace}
\makeatletter
\newcommand{\lyxaddress}[1]{
\par {\raggedright #1
\vspace{1.4em}
\noindent\par}
}

\usepackage[dvips]{graphicx}            
\usepackage{pslatex}

\usepackage[T1]{fontenc}
\usepackage{epsfig,url}
\usepackage[latin9]{inputenc}
\usepackage{geometry}
\usepackage{amsthm}
\usepackage{enumitem}
\usepackage{amstext}
\usepackage{setspace}

\usepackage{pstricks}
\usepackage{graphicx}
\usepackage{epsf}
\usepackage{epsfig}
\usepackage{feynarts}
\usepackage{babel}
\font \rus= wncyr10
\newcommand{\shuffle}{\, \hbox{\rus x} \,}

\usepackage{lipsum}

\let\OLDthebibliography\thebibliography
\renewcommand\thebibliography[1]{
  \OLDthebibliography{#1}
  \setlength{\parskip}{0pt}
  \setlength{\itemsep}{0pt plus 0.3ex}
}
\usepackage{amssymb}

\makeatother

\usepackage{babel}
\begin{document}

\title{MPL - a program for computations with iterated integrals on moduli
spaces of curves of genus zero \date{}}

\author{Christian Bogner}

\maketitle

\lyxaddress{\begin{center}
\emph{Institut f\"ur Physik, Humboldt-Universit\"at zu Berlin,
}\\
\emph{D - 10099 Berlin, Germany}
\par\end{center}}
\begin{abstract}
We introduce the computer program MPL for computations with homotopy
invariant iterated integrals on moduli spaces $\mathcal{M}_{0,n}$
of curves of genus 0 with $n$ ordered marked points. The program
is an implementation of the algorithms presented in \cite{BogBro2},
based on Maple. It includes the symbol map and procedures for the
analytic computation of period integrals on $\mathcal{M}_{0,n}.$
It supports the automated computation of a certain class of Feynman
integrals.

\pagebreak{}
\end{abstract}

\section{Introduction}

The use of polylogarithms and their generalizations has become standard
in the computation of Feynman integrals. A key-advantage of such classes
of functions is their double nature as nested sums and iterated integrals,
allowing for a choice between computational methods relying on summation
techniques and methods based on integral representations. 

While classical polylogarithms already appear in one-loop results,
the computation of higher-loop integrals often requires more general
classes of functions. The class of harmonic polylogarithms \cite{RemVer},
implemented in \cite{Mai1,Mai2,Ver,GehRem1}, serves for many computations,
particularly when combined with the method of computing the Feynman
integral by solving an associated differential equation \cite{Kot,Rem}.
Introducing the dependence on an additional parameter, harmonic polylogarithms
were extended%
\footnote{Further extensions of harmonic polylogarithms include functions of
\cite{Abletal,BirGloMar}.%
} in \cite{GehRem2,GehRem3}. If one continues along this line of extensions
to an arbitrary number of parameters, one arrives at a class of iterated
integrals known as hyperlogarithms, considered already in \cite{Poi}
and extensively discussed in \cite{Lap1,Lap2}. This class of iterated
integrals can be used to represent the multiple polylogarithms defined
in \cite{Gon2}. Their numerical evaluation was implemented in \cite{VolWei}.
An overview of properties of hyperlogarithms and their recent applications
in perturbative quantum field theory can be obtained from \cite{Pan1,Duh1}
and references therein. 

When writing the Feynman integral in terms of Feynman parameters,
it is often possible to successively integrate out these parameters,
building up the result as an iterated integral of an appropriate class.
By use of hyperlogarithms, this approach was systematized in \cite{Bro5}.
Finite integrals can be computed in this way, if certain polynomials
in the integrand satisfy the criterion of linear reducibility as defined
in \cite{Bro5} and later refined in \cite{Bro6}. This approach for
the computation of Feynman integrals in terms of hyperlogarithms was
fully implemented in the program \texttt{HyperInt} \cite{Pan2}.

Much of the recent progress on the mathematical understanding of Feynman
integral results arose from the idea to relate Feynman integrals to
period integrals, studied in algebraic geometry. In \cite{BloEsnKre}
this correspondence was made explicit by exhibiting a Feynman integral
whose result is the period of a motive associated to the first Symanzik
polynomial of the Feynman graph. An extensive exploration of periods
arising from $\phi^{4}$-theory was conducted in \cite{Sch}. Furthermore,
in a very general context, the coefficients of the Laurent series
of dimensionally regularized Feynman integrals are period integrals
\cite{BogWei} in the sense of \cite{KonZag}. Recent results at high
loop-order suggest, that even though not all periods arising from
Feynman graphs are contained in the set of multiple zeta values \cite{BroSch,BroDor},
the 'Feynman periods' seem to constitute a very particular subset
of periods with special properties \cite{PanSch}. 

Multiple zeta values do not only appear in Feynman integral computations
but also as periods of moduli spaces of curves of genus zero. In \cite{Bro1}
it was proven, that all periods of moduli spaces $\mathcal{M}_{0,n}$
of curves of genus zero with $n\geq3$ ordered, marked points are
multiple zeta values, as previously conjectured in \cite{GonMan}.
The proof uses the class of homotopy invariant iterated integrals
on these spaces and includes the statement, that this class is closed
under taking primitives. The program MPL, introduced in the following,
is based on this class of iterated integrals. 

In this article, we try to limit the discussion of the mathematical
background to aspects of direct importance for the use of the program
MPL. All further mathematical details may be obtained from \cite{BogBro2}
and \cite{Bro1}. Our program MPL is an implementation of the algorithms
presented in \cite{BogBro2}, which in turn are based on \cite{Bro1}.
The program can be seen as divided into two main parts. The first
part is dedicated to computations with iterated integrals on $\mathcal{M}_{0,n}$,
using the framework of so-called cubical coordinates and the corresponding
differential 1-forms. Apart from basic operations, such as the shuffle
product and the co-product of de-concatenation, the program provides
the construction of a basis for the vector space of these iterated
integrals by use of the so-called symbol map. It furthermore includes
procedures for the differentiation and the derivation of primitives
and exact limits at certain points. In particular, it allows for the
automated, analytical computation of a class of definite integrals
on $\mathcal{M}_{0,n}$. This class of integrals appears in many different
contexts, some of which were already pointed out in \cite{BogBro2}. 

The second part of the program is dedicated to the automated computation
of a certain class of Feynman integrals by the mentioned approach
of iteratively integrating out Feynman parameters. Here our strategy
is the following: For each Feynman parameter, we map the integrand
to differential 1-forms in cubical coordinates by an appropriate change
of variables. This reduces the problem to an integral on $\mathcal{M}_{0,n}$
of the mentioned class, which we compute by the methods of the first
part of the program. Then we map the result of this integration back
to an integral only in Feynman parameters and repeat these steps for
the remaining parameters. The program constructs and applies such
changes of variables, regarding the normalization conditions of both
representations at a tangential basepoint. 

The class of hyperlogarithms and the class of iterated integrals on
$\mathcal{M}_{0,n}$ are equivalently general, in the sense that both
can be used to express multiple polylogarithms and each other respectively.
In the context of the computation of Feynman integrals, each of these
classes of functions comes with its own advantages and drawbacks.
The differential 1-forms used to set-up the hyperlogarithms can be
defined by direct use of the polynomials, defining the singularities
of the given integrand in Feynman parameters. In this way, the integration
problem is easily formulated in terms of well adapted iterated integrals.
This advantage naturally comes with the inconvenience, that in principle,
for each new Feynman integral one works with a new set of iterated
integrals. 

Using iterated integrals on $\mathcal{M}_{0,n}$ instead, each integral
is expressed in terms of a finite basis of functions up to some weight
$w$ and number of variables $m.$ So in principle, if we consider
$w$ and $m$ large enough, we will always work with the same set
of iterated integrals. On the other hand, in order to use this convenient
framework, the given integrand has to be expressed in terms of the
particular differential 1-forms in cubical coordinates by a change
of variables. The need for such a change of variables introduces certain
restrictions on the integrand, to be made precise below. For cases
where these conditions are met, the integration procedure is fully
implemented in MPL. In other cases, it still may be possible for the
user to map the integrand to cubical coordinates 'by hand' and apply
MPL afterwards. Of course, one also has to expect cases of Feynman
integrals, where such changes of variables do not exist, including
the cases where multiple polylogarithms are not sufficient to express
the result. 

This article is structured as follows. Subsection \ref{sub:How-to-start}
contains information on how to obtain and start the program. Section
\ref{sec:Computing-with-iterated} introduces the framework of iterated
integrals on moduli spaces of curves of genus zero in terms of cubical
coordinates. Here we discuss basic operations, the symbol map and
the construction of the vectorspace of these integrals. In section
\ref{sec:Integration-over-cubical} we discuss the computation of
period integrals on the moduli spaces, introducing procedures for
taking primitives and certain limits. Section \ref{sec:Integration-over-Feynman}
adresses the problem of computing Feynman integrals. Here we specify
the conditions under which MPL can be applied and introduce procedures
to check these conditions and to compute the integral by iterative
integration over Feynman parameters. Section \ref{sec:Conclusions}
contains our conclusions. In appendix A we give a detailed example
of a Feynman integral computation with MPL and in appendix B we give
a very basic introduction to moduli spaces of curves of genus zero.

\subsection{How to start the program\label{sub:How-to-start}}

The latest version of \texttt{MPL} is available from the webpage

\texttt{http://cbogner.com/software/mpl/} \\
The entire program is obtained in one \texttt{txt}-file \texttt{MPLn\_m.txt},
where the integers $n$ and $m$ indicate the number of the version.
For example, the file of MPL version 1.0 is called \texttt{MPL1\_0.txt}.
After saving the file in the same directory with the Maple worksheet,
the program is started with
\begin{quote}
\texttt{>read(\textquotedbl{}MPL1\_0.txt\textquotedbl{}):}
\end{quote}
in the worksheet. While the most important procedures of the program
are introduced in this article, further technical details and examples
are provided in a user manual, which is also available from the above
webpage. MPL was written and tested with Maple 16.

For many applications, it is convenient to let all appearing multiple
zeta values be expressed in terms of an irreducible basis automatically.
We have used MPL with the file \texttt{mzv-1-12.txt} provided by \cite{Bigetal},
which serves for this purpose for multiple zeta values up to weight
12. In the Maple worksheet, it is started simply by 
\begin{quote}
\texttt{>read(\textquotedbl{}mzv-1-12.txt\textquotedbl{}):}
\end{quote}

\section{Computing with iterated integrals on moduli spaces of curves of genus
zero\label{sec:Computing-with-iterated}}

In this section, we define the differential 1-forms and iterated integrals
which our implementation is based on. We introduce procedures for
basic operations, the symbol and unshuffle maps and a construction
of the vectorspace of these iterated integrals.

\subsection{Iterated integrals\label{sub:Iterated-integrals}}

In MPL, every iterated integral is represented by an ordered sequence
of differential 1-forms. In order to motivate this notation, let us
briefly recall the concept of iterated integrals. For a general introduction
to the terminology we recommend \cite{Bro2}. 

We consider smooth differential 1-forms $\omega_{1},\,...,\,\omega_{k}$
on a smooth, complex manifold $M$ and a smooth path $\gamma:\,[0,\,1]\rightarrow M.$
An iterated integral along $\gamma$ is defined by 
\[
\int_{\gamma}\omega_{1}...\omega_{k}=\int_{0\leq t_{1}\leq...\leq t_{k}\leq1}\gamma^{\star}\left(\omega_{k}\right)\left(t_{1}\right)...\gamma^{\star}\left(\omega_{1}\right)\left(t_{k}\right)
\]
where $\gamma^{\star}\left(\omega_{i}\right)\left(t_{j}\right)$ denotes
the pull-back of $\omega_{i}$ along $\gamma$, evaluated at $t_{j}$
for $i,\, j=1,\,...,\, k.$ We consider the corresponding ordered
sequence of 1-forms $\omega_{1}\otimes\omega_{2}\otimes...\otimes\omega_{k}$
which we write in the so-called bar notation $\left[\omega_{1}|\omega_{2}|...|\omega_{k}\right].$
Note that in our convention, the iterated integration starts with
the rightmost 1-form and proceeds to the left in this sequence. In
general we will use the term \emph{iterated integral} for linear combinations
\[
I=\sum_{J=(i_{1},\,...,\, i_{k})}c_{J}\int_{\gamma}\omega_{i_{1}}...\omega_{i_{k}}
\]
 of such integrals and we call the corresponding linear combination
\begin{equation}
\omega=\sum_{J=(i_{1},\,...,\, i_{k})}c_{J}\left[\omega_{i_{1}}|...|\omega_{i_{k}}\right]\label{eq:linear combination of words}
\end{equation}
 of sequences the \emph{word} of $I.$

A famous theorem in \cite{Che} implies, that such an iterated integral
$I$ is homotopy invariant, if and only if its word $\omega$ satisfies
the so-called integrability condition 
\begin{equation}
\sum_{J=(i_{1},\,...,\, i_{k})}c_{J}\left(\left[\omega_{i_{1}}|...|\omega_{i_{j}}\wedge\omega_{i_{j+1}}|...|\omega_{i_{k}}\right]\right)=0\;\textrm{for all }1\leq j\leq k-1.\label{eq:Integrability Condition}
\end{equation}
A word $\omega$ satisfying this condition is called \emph{integrable}.
A homotopy invariant iterated integral $I$ depends on the homotopy
equivalence class of the path $\gamma$. Let $m$ be the dimension
of the manifold $M$ and let us write one of the end-points as $(x_{1},\, x_{2},...,\, x_{m})$
in some coordinates. By convention, we always choose the origin $(0,\,...,\,0)$
as the other end-point of $\gamma.$ As we will use differential 1-forms
with at most logarithmic poles below, one can show that all our iterated
integrals admit an expansion of the type 
\[
\sum_{J=(i_{1},\,...,\, i_{m})}f_{J}(x_{1},\,...,\, x_{m})\ln(x_{1})^{i_{1}}...\ln(x_{m})^{i_{m}}
\]
where the functions $f_{J}$ are analytic (thus in particular convergent)
at the origin. This expansion is used to regularize and to normalize
the iterated integrals: With respect to cubical coordinates $x_{1},\, x_{2},...,\, x_{m}$
introduced in the following subsection, the regularized value of the
function at the origin is defined to be the term $f_{(0,\,...,\,0)}(0,\,...,\,0)$,
i.e. the term obtained by setting all logarithms equal to zero in
the expansion. The normalization condition is 
\begin{equation}
f_{(0,\,...,\,0)}(0,\,...,\,0)=0.\label{eq:Normalization}
\end{equation}

Due to these conditions, every homotopy invariant iterated integral
$I$ is fully determined as a multivalued function of the variables
$x_{1},\, x_{2},...,\, x_{m}$ by the corresponding $\omega.$ 

Let us remark that in the physics literature, sometimes the term \emph{symbol}
is used as synonym for what we called integrable word here. In some
computations where only the differential behaviour of the corresponding
iterated integral is relevant, symbols are conveniently used without
further conditions with respect to the end-points of $\gamma.$ In
this case, symbols only represent the function up to contributions
of lower length. However, in this article, the above conditions fix
this ambiguity and every integrable word determines the corresponding
iterated integral entirely. Therefore we use the word $\omega$ to
denote the corresponding iterated integral $I.$

In our program, the bar notation is represented by the command \texttt{bar(...)}.
For example, a word $\left[a|b|c\right]$ is represented by 
\begin{quote}
\texttt{>bar(a, b, c):}
\end{quote}
in MPL and stands for the iterated integral obtained by integrating
over the 1-forms in the brackets from right to the left. Numerical
multiples are factored out automatically. For example we have:
\begin{quote}
\texttt{>bar(3{*}a, 2{*}b) + bar(0);}

$6\, bar(a,\, b)$ 
\end{quote}
Before specifying the differential 1-forms, let us discuss the multiplication
and co-multiplication of iterated integrals. Let $a=\left[a_{1}|...|a_{k}\right],\, b=\left[b_{1}|...|b_{m}\right]$
be two words of some differential 1-forms. By $a\sqcup b$ we denote
the concatenation: 
\[
a\sqcup b=\left[a_{1}|...|a_{k}|b_{1}|...|b_{m}\right].
\]
 The shuffle product $a\shuffle b$ is defined by 
\[
a\textrm{\shuffle}b=[a_{1}]\sqcup\left(\left[a_{2}|...|a_{k}\right]\shuffle b\right)+[b_{1}]\sqcup\left(a\shuffle\left[b_{2}|...|b_{m}\right]\right).
\]
If $\omega_{1}$ and $\omega_{2}$ are the words of iterated integrals
$I_{1}$ and $I_{2}$ respectively, then $\omega_{1}\shuffle\omega_{2}$
is the word of the product $I_{1}\cdot I_{2}.$ In MPL, the shuffle
product is implemented in the procedure \texttt{MPLShuffleProduct(a,b)}
where the two arguments are the words (in bar notation) to be multiplied
with each other. For example: 
\begin{quote}
\texttt{>MPLShuffleProduct(bar(u,v),bar(x,y)+7{*}bar(z));}\\
$7\, bar(u,v,z)+7\, bar(u,z,v)+7\, bar(z,u,v)+bar(u,v,x,y)+bar(u,x,v,y)+bar(u,x,y,v)+bar(x,u,v,y)+bar(x,u,y,v)+bar(x,y,u,v)$
\end{quote}
The de-concatenation co-product $\Delta$ , defined by 
\[
\Delta\left[a_{1}|a_{2}|...|a_{k}\right]=1\otimes\left[a_{1}|a_{2}|...|a_{k}\right]+\left[a_{1}\right]\otimes\left[a_{2}|...|a_{k}\right]+...+\left[a_{1}|...|a_{k}\right]\otimes1
\]
can be computed with the procedure \texttt{MPLCoproduct(...)}. Note
that MPL returns explicit tensor-products of words by use of \texttt{tens(...)}.

\subsection{Differential 1-forms in cubical coordinates}

The program MPL is based on the class of homotopy invariant iterated
integrals on moduli spaces $\mathcal{M}_{0,n}$ of curves with $n\geq3$
ordered, marked points. These spaces and this class of functions are
extensively studied in \cite{Bro1}. In appendix B, we give a very
basic introduction to the spaces $\mathcal{M}_{0,n}$. However, without
giving a full account on the underlying geometry here, the mentioned
class of iterated integrals can be specified as follows. For $m=n-3$
let us consider the set $\Omega_{m}$ of differential 1-forms, defined
by 
\[
\Omega_{m}=\left\{ \frac{dx_{1}}{x_{1}},...,\frac{dx_{m}}{x_{m}},\frac{d\left(\prod_{a\leq i\leq b}x_{i}\right)}{\prod_{a\leq i\leq b}x_{i}-1}\textrm{ for }1\leq a\leq b\leq m\right\} .
\]
Let $\mathcal{A}_{m}$ be the $\mathbb{Q}$-vectorspace spanned by
$\Omega_{m}.$ By the class of \emph{iterated integrals on $\mathcal{M}_{0,n}$}
we refer to the $\mathbb{Q}$-vectorspace $V\left(\Omega_{m}\right)$
of homotopy invariant iterated integrals of differential 1-forms in
$\mathcal{A}_{m},$ regularized and normalized by the conditions introduced
in section \ref{sub:Iterated-integrals}. 

The variables $x_{1},\,...,\, x_{m}$ in the above 1-forms are called
\emph{cubical coordinates} in this context. We will sometimes refer
to $x_{m}$ as the \emph{last} of the cubical coordinates. In the
Maple worksheet, the set of cubical coordinates has to be declared
before many of the computations below. With the command \texttt{MPLCoordinates(letter,
m)} one declares $m$ cubical coordinates, where the first argument
is used to construct the variable names. For example after the command
\begin{quote}
\texttt{>MPLCoordinates(y, 3):}
\end{quote}
we can compute with \texttt{y{[}1{]}}, \texttt{y{[}2{]}}, \texttt{y{[}3{]}}. 

Following \cite{BogBro2}, let us furthermore introduce the auxiliary
sets of differential 1-forms 
\begin{eqnarray*}
\bar{\Omega}_{m}^{F} & = & \left\{ \frac{dx_{m}}{x_{m}},\frac{\left(\prod_{a\leq i\leq m-1}x_{i}\right)dx_{m}}{\prod_{a\leq i\leq m}x_{i}-1}\textrm{ for }1\leq a\leq m\right\} ,\\
\Omega_{m}^{F} & = & \left\{ \frac{dx_{m}}{x_{m}},\frac{d\left(\prod_{a\leq i\leq m}x_{i}\right)}{\prod_{a\leq i\leq m}x_{i}-1}\textrm{ for }1\leq a\leq m\right\} ,
\end{eqnarray*}
noting that $\Omega_{m}=\Omega_{m}^{F}\cup\Omega_{m-1}.$ We define
$\mathcal{\bar{A}}_{m}^{F},\,\mathcal{A}_{m}^{F}$ to be the $\mathbb{Q}$-vectorspaces
of differential 1-forms, spanned by the bases $\bar{\Omega}_{m}^{F},\,\Omega_{m}^{F}$
respectively. An isomorphism between these spaces is given by 
\begin{eqnarray*}
\lambda_{m}:\,\mathcal{\bar{A}}_{m}^{F} & \rightarrow & \mathcal{A}_{m}^{F},\\
\frac{dx_{m}}{x_{m}} & \mapsto & \frac{dx_{m}}{x_{m}},\\
\frac{\left(\prod_{a\leq i\leq m-1}x_{i}\right)dx_{m}}{\prod_{a\leq i\leq m}x_{i}-1} & \mapsto & \frac{d\left(\prod_{a\leq i\leq m}x_{i}\right)}{\prod_{a\leq i\leq m}x_{i}-1}.
\end{eqnarray*}

The $\mathbb{Q}$-vectorspace $V\left(\bar{\Omega}_{m}^{F}\right)$
of iterated integrals with all 1-forms in $\mathcal{\bar{A}}_{m}^{F}$
plays an auxiliary role in some computations. While the iterated integrals
in $V\left(\Omega_{m}\right)$ are functions of $m$ variables on
$\mathcal{M}_{0,n}$, $V\left(\bar{\Omega}_{m}^{F}\right)$ is a space
of hyperlogarithms, being functions of the one variable $x_{m}$ on
a fiber over $\mathcal{M}_{0,n-1},$ with the $x_{1},\,...,\, x_{m-1}$
considered fixed. Note that every word in $\mathcal{\bar{A}}_{m}^{F}$
is integrable by construction of $\bar{\Omega}_{m}^{F}$, but words
in $\mathcal{A}_{m}$ can fail the integrability condition. An explicit
construction of all integrable words in $\mathcal{A}_{m}$, implying
the construction of $V\left(\Omega_{m}\right)$, is discussed in section
\ref{sub:The-symbol-map}. 

Let us mention that the above differential 1-forms satisfy quadratic
relations due to Arnold \cite{Arn}, of the form 
\begin{equation}
\omega_{i}\wedge\omega_{j}=\sum_{k}\alpha_{k}\wedge\omega_{k},\label{eq:Arnold equation}
\end{equation}
with $\omega_{i}\in\Omega_{m}^{F}$ $\alpha_{i}\in\Omega_{m-1}$.
These Arnold equations are explicitely given in \cite{BogBro2} and
are of internal importance for our algorithms. By the command \texttt{MPLArnoldEquation(...)},
they are also available for the user. We refer to the manual for details.

\subsection{Differentiation}

There are several notions of differentiation on $V\left(\Omega_{m}\right)$
and $V\left(\bar{\Omega}_{m}^{F}\right).$ For both spaces, the differentiation
$d$ with respect to the end-point of the path is simply the truncation
of the leftmost 1-form:
\begin{eqnarray*}
d:\, V\left(\Omega_{m}\right) & \rightarrow & \mathcal{A}_{m}\otimes V\left(\Omega_{m}\right),\\
\sum_{J=(i_{1},\,...,\, i_{k})}c_{J}\left[\omega_{i_{1}}|...|\omega_{i_{k}}\right] & \mapsto & \sum_{J=(i_{1},\,...,\, i_{k})}c_{J}\omega_{i_{1}}\otimes\left[\omega_{i_{2}}|...|\omega_{i_{k}}\right].
\end{eqnarray*}
This operation is well-known from the literature on hyperlogarithms.
It is implemented in the procedure \texttt{MPLd(...)}. 

A\emph{ connection} is a linear map
\[
\nabla:\, V\left(\bar{\Omega}_{m}^{F}\right)\rightarrow\mathcal{A}_{m-1}\otimes V\left(\bar{\Omega}_{m}^{F}\right)
\]
satisfying the Leibniz rule. In our framework, it can be constructed
\cite{BogBro2} by firstly applying the map $\lambda_{m}$ to a word
in $\mathcal{\bar{A}}_{m}^{F},$ then applying an operator $D$ defined
by 
\[
D\left[\omega_{1}|...|\omega_{k}\right]=(-1)^{k}\sum_{i=1}^{k-1}\left[\omega_{1}|...|\omega_{i}\wedge\omega_{i+1}|\omega_{k}\right],
\]
then expressing the wedge products on the right-hand in the form $\sum_{k}\alpha_{k}\wedge\omega_{k}$
with $\omega_{i}\in\Omega_{m}^{F}$ $\alpha_{i}\in\Omega_{m-1}$ by
use of the Arnold equations eq. \ref{eq:Arnold equation} and finally
pulling out all $\alpha_{i}$ to the left. With the help of the connection
$\nabla,$ the \emph{total connection} 
\[
\nabla_{T}:\, V\left(\bar{\Omega}_{m}^{F}\right)\rightarrow\mathcal{A}_{m}\otimes V\left(\bar{\Omega}_{m}^{F}\right)
\]
is defined by 
\[
\nabla_{T}=d-\nabla.
\]
It is implemented in the procedure \texttt{MPLTotalConnection(...)}.
We refer to the manual for examples.

\subsection{The symbol map and the construction of $V\left(\Omega_{m}\right)$\label{sub:The-symbol-map}}

As not every word in $\mathcal{A}_{m}$ is integrable, the construction
of the vectorspace $V\left(\Omega_{m}\right)$ is not trivial. For
example let $m=3$ and consider the words 
\[
\omega_{1}=\left[\frac{dx_{3}}{x_{3}}+\frac{dx_{2}}{x_{2}}|\frac{d\left(x_{2}x_{3}\right)}{x_{2}x_{3}-1}\right]\textrm{ and }\omega_{2}=\left[\frac{dx_{3}}{x_{3}}|\frac{d\left(x_{2}x_{3}\right)}{x_{2}x_{3}-1}\right].
\]
The word $\omega_{1}$ is integrable, but $\omega_{2}$ is not. MPL
does not prohibit the use of non-integrable words, and many operations
can be applied to $\omega_{2}$ as well as to $\omega_{1}.$ However,
$\omega_{2}$ does not represent an iterated integral in our framework.
In many applications, the user does not need to worry about this point.
If MPL is used for the computation of an integral, the program returns
the result in terms of iterated integrals which are homotopy invariant
by construction. However, for some applications, it may be useful
to have an explicit basis for $V\left(\Omega_{m}\right)$ up to a
certain weight at hand%
\footnote{The construction of certain subspaces of integrable words played an
important role e.g. in the recent computations \cite{DixDruHipPen,DixDruDuhPen,DruPapSpr}.%
}. 

The construction of this basis is facilitated by the so-called symbol
map

\[
\Psi:\, V\left(\bar{\Omega}_{m}^{F}\right)\rightarrow V\left(\Omega_{m}\right).
\]
It is the unique linear map satisfying 
\[
\left(id\otimes\Psi\right)\circ\nabla_{T}=d\circ\Psi.
\]
This map was explicitely constructed in \cite{BogBro1,BogBro2} and
is related to the constructions of the symbol in \cite{DuhGanRho,Gon1,Gonetal}.
It is implemented in \texttt{MPLSymbolMap(...)}.

\subsubsection*{Example:}

We apply the symbol map $\Psi$ to the hyperlogarithm $\left[\frac{dx_{3}}{x_{3}}|\frac{x_{2}d\left(x_{3}\right)}{1-x_{2}x_{3}}\right]\in V\left(\bar{\Omega}_{m}^{F}\right):$
\begin{quote}
\texttt{>MPLCoordinates(x,3):}~\\
\texttt{>MPLSymbolMap(bar(d(x{[}3{]})/(x{[}3{]}), x{[}2{]}{*}d(x{[}3{]})/(1-x{[}2{]}{*}x{[}3{]})));}~\\
\texttt{$bar(d(x[3])/x[3],(x[3]*d(x[2])+x[2]*d(x[3]))/(1-x[2]*x[3]))+bar(d(x[2])/x[2],(x[3]*d(x[2])+x[2]*d(x[3]))/(1-x[2]*x[3]))$}
\end{quote}
Slightly simplifying this output, we obtain \texttt{
\begin{equation}
\Psi\left(\left[\frac{dx_{3}}{x_{3}}|\frac{x_{2}d\left(x_{3}\right)}{1-x_{2}x_{3}}\right]\right)=\left[\frac{dx_{3}}{x_{3}}+\frac{dx_{2}}{x_{2}}|\frac{d\left(x_{2}x_{3}\right)}{1-x_{2}x_{3}}\right]\in V\left(\Omega_{3}\right).\label{eq:Symbol example}
\end{equation}
}

Let $\bar{B}_{m,w}$ be the set of iterated integrals with words $\omega=\left[\omega_{1}|...|\omega_{k}\right]$
with $k\leq w$ and with all 1-forms in $\bar{\Omega}_{m}^{F}.$ This
set $\bar{B}_{m,w}$ is a basis for the subspace $V_{w}\left(\bar{\Omega}_{m}^{F}\right)\subset V\left(\bar{\Omega}_{m}^{F}\right)$
of iterated integrals whose words in $\mathcal{\bar{A}}_{m}^{F}$
are of length $k\leq w.$ Let $B_{m,w}$ be the desired basis of the
subspace $V_{w}\left(\Omega_{m}\right)\subset V\left(\Omega_{m}\right)$
of words of length $k\leq w.$ A theorem of \cite{Bro1} states the
existence of an isomorphism of algebras
\begin{equation}
V\left(\Omega_{m}\right)\cong V\left(\Omega_{m-1}\right)\otimes V\left(\bar{\Omega}_{m}^{F}\right).\label{eq:Decomposition}
\end{equation}
As a consequence, $B_{m,w}$ is the set of all iterated integrals
whose words $\omega$ are of length $k\leq w$ and are obtained as
products $\omega=\eta\shuffle\Psi\left(\xi\right)$ with $\eta\in B_{m-1,w}$
and $\xi\in\bar{B}_{m,w}.$ We can formulate this construction by
the map 
\begin{equation}
\mu\left(id\otimes\Psi\right):\, V\left(\Omega_{m-1}\right)\otimes V\left(\bar{\Omega}_{m}^{F}\right)\rightarrow V\left(\Omega_{m}\right)\label{eq:Shuffle and Symbol}
\end{equation}
where $\mu$ denotes multiplication. This provides a recursive construction
of the basis $B_{m,w}$. The procedure \texttt{MPLBasis(letter,m,w)}
returns $B_{m,w}$ for words up to weight \texttt{w} in variables
named by the first argument.

\subsubsection*{Example: }
\begin{quote}
\texttt{>MPLBasis(y,2,2);
\[
\left[\left[{\it bar}\left({\frac{d\left(y_{{2}}\right)}{y_{{2}}}}\right),{\it bar}\left({\frac{d\left(y_{{2}}\right)}{1-y_{{2}}}}\right),{\it bar}\left({\frac{y_{{2}}d\left(y_{{1}}\right)+y_{{1}}d\left(y_{{2}}\right)}{1-y_{{1}}y_{{2}}}}\right),{\it bar}\left({\frac{d\left(y_{{1}}\right)}{y_{{1}}}}\right),{\it bar}\left({\frac{d\left(y_{{1}}\right)}{1-y_{{1}}}}\right)\right],\right.
\]
\[
\left[{\it bar}\left({\frac{d\left(y_{{2}}\right)}{y_{{2}}}},{\frac{d\left(y_{{2}}\right)}{y_{{2}}}}\right),{\it bar}\left({\frac{d\left(y_{{2}}\right)}{y_{{2}}}},{\frac{d\left(y_{{2}}\right)}{1-y_{{2}}}}\right),{\it bar}\left({\frac{d\left(y_{{2}}\right)}{1-y_{{2}}}},{\frac{d\left(y_{{2}}\right)}{y_{{2}}}}\right),\right.
\]
\[
{\it bar}\left({\frac{d\left(y_{{2}}\right)}{y_{{2}}}},{\frac{y_{{2}}d\left(y_{{1}}\right)+y_{{1}}d\left(y_{{2}}\right)}{1-y_{{1}}y_{{2}}}}\right)+{\it bar}\left({\frac{d\left(y_{{1}}\right)}{y_{{1}}}},{\frac{y_{{2}}d\left(y_{{1}}\right)+y_{{1}}d\left(y_{{2}}\right)}{1-y_{{1}}y_{{2}}}}\right),
\]
\[
{\it bar}\left({\frac{y_{{2}}d\left(y_{{1}}\right)+y_{{1}}d\left(y_{{2}}\right)}{1-y_{{1}}y_{{2}}}},{\frac{d\left(y_{{2}}\right)}{y_{{2}}}}\right)-{\it bar}\left({\frac{d\left(y_{{1}}\right)}{y_{{1}}}},{\frac{y_{{2}}d\left(y_{{1}}\right)+y_{{1}}d\left(y_{{2}}\right)}{1-y_{{1}}y_{{2}}}}\right),{\it bar}\left({\frac{d\left(y_{{2}}\right)}{1-y_{{2}}}},{\frac{d\left(y_{{2}}\right)}{1-y_{{2}}}}\right),
\]
\[
{\it bar}\left({\frac{d\left(y_{{2}}\right)}{1-y_{{2}}}},{\frac{y_{{2}}d\left(y_{{1}}\right)+y_{{1}}d\left(y_{{2}}\right)}{1-y_{{1}}y_{{2}}}}\right)-{\it bar}\left({\frac{d\left(y_{{1}}\right)}{1-y_{{1}}}},{\frac{y_{{2}}d\left(y_{{1}}\right)+y_{{1}}d\left(y_{{2}}\right)}{1-y_{{1}}y_{{2}}}}\right)+{\it bar}\left({\frac{d\left(y_{{1}}\right)}{1-y_{{1}}}},{\frac{d\left(y_{{2}}\right)}{1-y_{{2}}}}\right)
\]
\[
-{\it bar}\left({\frac{d\left(y_{{1}}\right)}{y_{{1}}}},{\frac{y_{{2}}d\left(y_{{1}}\right)+y_{{1}}d\left(y_{{2}}\right)}{1-y_{{1}}y_{{2}}}}\right),{\it bar}\left({\frac{y_{{2}}d\left(y_{{1}}\right)+y_{{1}}d\left(y_{{2}}\right)}{1-y_{{1}}y_{{2}}}},{\frac{d\left(y_{{2}}\right)}{1-y_{{2}}}}\right)
\]
\[
+{\it bar}\left({\frac{d\left(y_{{1}}\right)}{1-y_{{1}}}},{\frac{y_{{2}}d\left(y_{{1}}\right)+y_{{1}}d\left(y_{{2}}\right)}{1-y_{{1}}y_{{2}}}}\right)-{\it bar}\left({\frac{d\left(y_{{1}}\right)}{1-y_{{1}}}},{\frac{d\left(y_{{2}}\right)}{1-y_{{2}}}}\right)
\]
\[
+{\it bar}\left({\frac{d\left(y_{{1}}\right)}{y_{{1}}}},{\frac{y_{{2}}d\left(y_{{1}}\right)+y_{{1}}d\left(y_{{2}}\right)}{1-y_{{1}}y_{{2}}}}\right),{\it bar}\left({\frac{y_{{2}}d\left(y_{{1}}\right)+y_{{1}}d\left(y_{{2}}\right)}{1-y_{{1}}y_{{2}}}},{\frac{y_{{2}}d\left(y_{{1}}\right)+y_{{1}}d\left(y_{{2}}\right)}{1-y_{{1}}y_{{2}}}}\right),
\]
\[
{\it bar}\left({\frac{d\left(y_{{1}}\right)}{y_{{1}}}},{\frac{d\left(y_{{2}}\right)}{y_{{2}}}}\right)+{\it bar}\left({\frac{d\left(y_{{2}}\right)}{y_{{2}}}},{\frac{d\left(y_{{1}}\right)}{y_{{1}}}}\right),{\it bar}\left({\frac{d\left(y_{{1}}\right)}{y_{{1}}}},{\frac{d\left(y_{{2}}\right)}{1-y_{{2}}}}\right)
\]
\[
+{\it bar}\left({\frac{d\left(y_{{2}}\right)}{1-y_{{2}}}},{\frac{d\left(y_{{1}}\right)}{y_{{1}}}}\right),{\it bar}\left({\frac{d\left(y_{{1}}\right)}{y_{{1}}}},{\frac{y_{{2}}d\left(y_{{1}}\right)+y_{{1}}d\left(y_{{2}}\right)}{1-y_{{1}}y_{{2}}}}\right)+{\it bar}\left({\frac{y_{{2}}d\left(y_{{1}}\right)+y_{{1}}d\left(y_{{2}}\right)}{1-y_{{1}}y_{{2}}}},{\frac{d\left(y_{{1}}\right)}{y_{{1}}}}\right),
\]
\[
{\it bar}\left({\frac{d\left(y_{{1}}\right)}{1-y_{{1}}}},{\frac{d\left(y_{{2}}\right)}{y_{{2}}}}\right)+{\it bar}\left({\frac{d\left(y_{{2}}\right)}{y_{{2}}}},{\frac{d\left(y_{{1}}\right)}{1-y_{{1}}}}\right),{\it bar}\left({\frac{d\left(y_{{1}}\right)}{1-y_{{1}}}},{\frac{d\left(y_{{2}}\right)}{1-y_{{2}}}}\right)+{\it bar}\left({\frac{d\left(y_{{2}}\right)}{1-y_{{2}}}},{\frac{d\left(y_{{1}}\right)}{1-y_{{1}}}}\right),
\]
\[
{\it bar}\left({\frac{d\left(y_{{1}}\right)}{1-y_{{1}}}},{\frac{y_{{2}}d\left(y_{{1}}\right)+y_{{1}}d\left(y_{{2}}\right)}{1-y_{{1}}y_{{2}}}}\right)+{\it bar}\left({\frac{y_{{2}}d\left(y_{{1}}\right)+y_{{1}}d\left(y_{{2}}\right)}{1-y_{{1}}y_{{2}}}},{\frac{d\left(y_{{1}}\right)}{1-y_{{1}}}}\right),{\it bar}\left({\frac{d\left(y_{{1}}\right)}{y_{{1}}}},{\frac{d\left(y_{{1}}\right)}{y_{{1}}}}\right),
\]
\[
\left.\left.{\it bar}\left({\frac{d\left(y_{{1}}\right)}{y_{{1}}}},{\frac{d\left(y_{{1}}\right)}{1-y_{{1}}}}\right),{\it bar}\left({\frac{d\left(y_{{1}}\right)}{1-y_{{1}}}},{\frac{d\left(y_{{1}}\right)}{y_{{1}}}}\right),{\it bar}\left({\frac{d\left(y_{{1}}\right)}{1-y_{{1}}}},{\frac{d\left(y_{{1}}\right)}{1-y_{{1}}}}\right)\right]\right]
\]
}
\end{quote}
This result is a basis for iterated integrals in $V\left(\Omega_{2}\right)$
in variables $y_{1},\, y_{2},\, y_{3},\, y_{4}$ up to weight 2 as
sequence of two lists $S=\left[S_{1},S_{2}\right].$ The list $S_{1}$
contains the basis for weight 1 and $S_{2}$ for weight 2.

\subsection{The unshuffle map and hyperlogarithms\label{sub:The-unshuffle-map}}

There is an explicit construction \cite{BogBro2} of the inverse of
the map of eq. \ref{eq:Shuffle and Symbol} called the unshuffle map:
\begin{equation}
\Phi:\, V\left(\Omega_{m}\right)\rightarrow V\left(\Omega_{m-1}\right)\otimes V\left(\bar{\Omega}_{m}^{F}\right).\label{eq:Unshuffle}
\end{equation}
It is available by the procedure \texttt{MPLUnshuffle(f,var)}, which
decomposes a function \texttt{f}, such that the right component of
tensor-product in eq. \ref{eq:Unshuffle} is a hyperlogarithm in the
variable \texttt{var. }The variable \texttt{var} has to be the last
of the declared cubical coordinates.

\subsubsection*{Example:}

Consider the function 
\[
f=\left[2\frac{dx_{2}}{x_{2}}+\frac{dx_{3}}{x_{3}}|\frac{dx_{2}}{x_{2}}|\frac{d\left(x_{2}x_{3}\right)}{1-x_{2}x_{3}}\right]+\left[\frac{dx_{2}}{x_{2}}|\frac{dx_{3}}{x_{3}}|\frac{d\left(x_{2}x_{3}\right)}{1-x_{2}x_{3}}\right]+\left[\frac{dx_{2}}{x_{2}}+\frac{dx_{3}}{x_{3}}|\frac{d\left(x_{2}x_{3}\right)}{1-x_{2}x_{3}}|\frac{dx_{2}}{x_{2}}\right]\in V\left(\Omega_{3}\right).
\]
We apply \texttt{MPLUnshuffle }to $f$ with respect to variable $x_{3}$:
\begin{quote}
\texttt{>MPLCoordinates(x,3):}~\\
\texttt{>MPLUnshuffle(f,x{[}3{]});}~\\
\texttt{$tens(bar(d(x[2])/x[2]),bar(d(x[3])/x[3],x[2]d(x[3])/(1-x[2]x[3])))$}
\end{quote}
This output is understood as
\[
\left[\frac{dx_{2}}{x_{2}}\right]\otimes\left[\frac{dx_{3}}{x_{3}}|\frac{x_{2}d\left(x_{3}\right)}{1-x_{2}x_{3}}\right].
\]
We see that the left part of this tensor product is a function in
$V\left(\Omega_{2}\right)$ (in this simple example, it even belongs
to $V\left(\bar{\Omega}_{2}^{F}\right)$), while the right part is
a hyperlogarithm in $V\left(\bar{\Omega}_{3}^{F}\right).$ Hence,
as easily confirmed using eq. \ref{eq:Symbol example}, the function
$f$ can be constructed as 
\[
f=\left[\frac{dx_{2}}{x_{2}}\right]\shuffle\Psi\left(\left[\frac{dx_{3}}{x_{3}}|\frac{x_{2}d\left(x_{3}\right)}{1-x_{2}x_{3}}\right]\right).
\]

In the same recursive way as the shuffle map $\Psi$ can be used to
construct all iterated integrals in $V\left(\Omega_{m}\right)$ from
hyperlogarithms in $V\left(\bar{\Omega}_{k}^{F}\right)$ with $k\leq m,$
the recursive application of the unshuffle map $\Phi$ to the left
part of the tensor product in eq. \ref{eq:Unshuffle} decomposes every
function in $V\left(\Omega_{m}\right)$ as a product of such hyperlogarithms.
In this way, one can express all functions of $V\left(\Omega_{m}\right)$
in terms of hyperlogarithms and vice versa.

\section{Integration over cubical coordinates\label{sec:Integration-over-cubical}}

In this section, we introduce procedures for the computation of primitives
and certain limits of iterated integrals in $V\left(\Omega_{m}\right).$
A further command which combines these computations serves for the
analytical computation of certain definite integrals on $\mathcal{M}_{0,n}.$
These procedures will also be the backbone of the Feynman integral
computations of section \ref{sec:Integration-over-Feynman}.

\subsection{Primitives\label{sub:Primitives}}

It is well-known and implied by the definition of hyperlogarithms,
that the computation of the primitive of a hyperlogarithm with respect
to some differential 1-form $\omega_{0}$ is simply the left-concatenation
of this 1-form to the corresponding word. In our set-up this is a
map 
\begin{eqnarray}
\bar{\mathcal{A}}_{m}^{F}\otimes V\left(\bar{\Omega}_{m}^{F}\right) & \rightarrow & V\left(\bar{\Omega}_{m}^{F}\right),\nonumber \\
\omega_{0}\otimes\left[\omega_{1}|...|\omega_{k}\right] & \mapsto & \int\omega_{0}\otimes\left[\omega_{1}|...|\omega_{k}\right]=\left[\omega_{0}|\omega_{1}|...|\omega_{k}\right].\label{eq:primitive by simple concatenation}
\end{eqnarray}
As every word in $\bar{\mathcal{A}}_{m}^{F}$ is integrable, every
primitive obtained in this way is trivially homotopy invariant. 

However, the computation of a primitive of a function $f\in V\left(\Omega_{m}\right)$
has to be different from simple left-concatenation in general. This
can already be understood from the simple fact that there are 1-forms
$\omega_{0}$ and integrable words $\omega$ with letters in $\Omega_{m},$
such that the concatenation $\omega_{0}\sqcup\omega$ is not integrable.
However, the primitive of a function in $V\left(\Omega_{m}\right)$
with respect to a 1-form in $\bar{\mathcal{A}}_{m}^{F}$ is given
by a map

\begin{eqnarray*}
\bar{\mathcal{A}}_{m}^{F}\otimes V\left(\Omega_{m}\right) & \rightarrow & V\left(\Omega_{m}\right),\\
\omega_{0}\otimes\left[\omega_{1}|...|\omega_{k}\right] & \mapsto & \int\omega_{0}\otimes\left[\omega_{1}|...|\omega_{k}\right]
\end{eqnarray*}
whose existence is implied by a theorem of \cite{Bro1}. 

In principle, an explicit computation of primitives could start from
a decomposition into hyperlogarithms as in section \ref{sub:The-unshuffle-map},
apply simple concatenation as in eq. \ref{eq:primitive by simple concatenation}
and then map back to $V\left(\Omega_{m}\right)$ by use of the symbol
map. However, a more efficient algorithm, avoiding the decomposition
into hyperlogarithms, was presented in \cite{BogBro2}. This algorithm
is implemented in MPL. For a 1-form \texttt{a} in $\bar{\mathcal{A}}_{m}^{F}$
and a function \texttt{f} in $V\left(\Omega_{m}\right)$ whose maximal
weight is \texttt{w}, the procedure \texttt{MPLPrimitive(a, f, w)}
computes the primitive $\int a\otimes f\in V\left(\Omega_{m}\right).$

\subsubsection*{Example:}

Let us consider the function 
\[
f=\left[\frac{dx_{2}}{x_{2}}+\frac{dx_{3}}{x_{3}}|\frac{d\left(x_{2}x_{3}\right)}{1-x_{2}x_{3}}\right]+17\left[\frac{dx_{3}}{1-x_{3}}\right]\in V\left(\Omega_{3}\right)
\]
whose maximal weight is $w=2,$ and the 1-form $a=\frac{dx_{3}}{1-x_{3}}\in\bar{\mathcal{A}}_{3}^{F}.$
By use of the commands 
\begin{quote}
\texttt{>MPLCoordinates(x,3):}

\texttt{>MPLPrimitive(a,f,2);}
\end{quote}
we obtain 
\begin{quote}
\begin{eqnarray*}
\int a\otimes f & = & \left[\frac{dx_{3}}{1-x_{3}}|\frac{dx_{3}}{x_{3}}|\frac{d\left(x_{2}x_{3}\right)}{1-x_{2}x_{3}}\right]+\left[\frac{dx_{3}}{1-x_{3}}-\frac{dx_{2}}{x_{2}}|\frac{dx_{2}}{x_{2}}|\frac{d\left(x_{2}x_{3}\right)}{1-x_{2}x_{3}}\right]+\left[\frac{dx_{2}}{x_{2}}|\frac{dx_{3}}{1-x_{3}}|\frac{d\left(x_{2}x_{3}\right)}{1-x_{2}x_{3}}\right]\\
 &  & +\left[\frac{dx_{2}}{x_{2}}|\frac{dx_{2}}{1-x_{2}}|\frac{dx_{3}}{1-x_{3}}-\frac{d\left(x_{2}x_{3}\right)}{1-x_{2}x_{3}}\right]+17\left[\frac{dx_{3}}{1-x_{3}}|\frac{dx_{3}}{1-x_{3}}\right]\in V\left(\Omega_{3}\right).
\end{eqnarray*}

\end{quote}

\subsection{Limits\label{sub:Limits}}

MPL can be used to take limits of functions $f\in V\left(\Omega_{m}\right)$
at $x_{k}=u$ where $u\in\{0,\,1\}$ and where $x_{k}$ is any of
the cubical coordinates $x_{1},\,...,\, x_{m}.$ Let $\mathcal{Z}$
be the $\mathbb{Q}$-vectorspace of multiple zeta values. It was proven
in \cite{Bro1} that for every $f\in V\left(\Omega_{m}\right)$ the
limits $\lim_{x_{k}\rightarrow u}f$ are $\mathcal{Z}$-linear combinations
of functions in $V\left(\Omega_{m-1}\right).$ 

If the limit is computed with respect to the last variable $x_{m},$
the limit is readily expressed in terms of 1-forms in $\Omega_{m-1}.$
In the case of $k<m,$ the limit $\lim_{x_{k}\rightarrow u}f$ may
involve iterated integrals in differential 1-forms of the type $\frac{d(x_{i}x_{i+1}...\hat{x}_{k}...x_{j})}{1-x_{i}x_{i+1}...\hat{x}_{k}...x_{j}},$
where the hat indicates the missing variable $x_{k}$ in the products
of consecutive coordinates. Here one has to re-name the coordinates
$(x_{k+1},\,...,\, x_{m})\mapsto(x_{k},\,...,\, x_{m-1}),$ such that
these 1-forms belong to $\Omega_{m-1}$ and the iterated integrals
are recognized to belong to $V\left(\Omega_{m-1}\right)$ by MPL. 

The limits are computed by expansion of the given $f\in V\left(\Omega_{m}\right)$
as a series in $x_{k}=u,$ and evaluation of the coefficient of $\ln(x-u)^{0}$
(according to the regularization condition of section \ref{sub:Iterated-integrals}).
The expansion makes use of the decomposition eq. \ref{eq:Decomposition}
such that the problem is internally reduced to the computation of
regularized limits of functions in $V\left(\Omega_{1}\right).$ The
analytical solution to the latter problem is known and leads to elements
of $\mathcal{Z}$ (see \cite{Bro3}). The computation does not involve
any numerical approximation. The command \texttt{MPLLimit(f,x{[}k{]},u)}
returns the limit of $f\in V\left(\Omega_{m}\right)$ at $x_{k}=u$
with $u\in\{0,\,1\}.$

\subsubsection*{Example: }

Consider the function 
\begin{equation}
f=\Psi\left(\left[\frac{dx_{2}}{1-x_{2}}|\frac{x_{1}dx_{2}}{1-x_{1}x_{2}}\right]\right)=\left[\frac{dx_{2}}{1-x_{2}}-\frac{dx_{1}}{1-x_{1}}-\frac{dx_{1}}{x_{1}}|\frac{d\left(x_{1}x_{2}\right)}{1-x_{1}x_{2}}\right]+\left[\frac{dx_{1}}{1-x_{1}}|\frac{dx_{2}}{1-x_{2}}\right]\in V\left(\Omega_{2}\right).\label{eq:example limit}
\end{equation}
By use of the commands 
\begin{quote}
\texttt{>MPLCoordinates(x,2):}~\\
\texttt{>MPLLimit(f,x{[}2{]},1);}
\end{quote}
we obtain the limit 
\[
\lim_{x_{2}\rightarrow1}f=-\left[\frac{dx_{1}}{1-x_{1}}+\frac{dx_{1}}{x_{1}}|\frac{dx_{1}}{1-x_{1}}\right]=-\frac{1}{2}\ln(1-x_{1})^{2}-\textrm{Li}_{2}(x_{1})\in V\left(\Omega_{2}\right)
\]
where $\textrm{Li}_{2}$ denotes the classical dilogarithm. 

In some applications it is necessary to compute several consecutive
limits 
\[
\lim_{x_{k_{1}}\rightarrow u_{1}}\lim_{x_{k_{2}}\rightarrow u_{2}}...\lim_{x_{k_{l}}\rightarrow u_{l}}f
\]
with $\{k_{1},\,...,\, k_{l}\}\subseteq\{1,\,...,\, m\}$ and $u_{i}\in\{0,\,1\}$
for $i\in\{1,\,...,\, l\}.$ Such computations are facilitated by
the procedure \texttt{MPLMultipleLimit(f,S)} where the first argument
is a function in $V\left(\Omega_{m}\right)$ and the second argument
is an ordered list \texttt{S={[}...{]}} of equations $x_{k_{l}}=u_{l},\, x_{k_{l-1}}=u_{l-1},\,...,\, x_{k_{1}}=u_{1}$
which define the limits. The order in which the limits are computed
is from left to right in this list. 

It is important to note that in such multiple limits, the order matters.
For example, for $f$ defined in eq. \ref{eq:example limit}, we have
\begin{eqnarray*}
\lim_{x_{1}\rightarrow1}\lim_{x_{2}\rightarrow1}f & = & -\zeta(2),\\
\lim_{x_{2}\rightarrow1}\lim_{x_{1}\rightarrow1}f & = & 0.
\end{eqnarray*}
In both computations, we approach the point $(x_{1},\, x_{2})=(1,\,1)$.
In the first line, we evaluate $f$ on the boundary $L_{1}=\left\{ (x_{1},\,1)|0\leq x_{1}\leq1\right\} $
at first and then approach the point $(1,\,1)$ along this line. In
the second computation, we move to the boundary $L_{2}=\left\{ (1,\, x_{2})|0\leq x_{2}\leq1\right\} $
at first and approach the point $(1,\,1)$ from there. The difference
between both results comes from the fact, that the point $(1,\,1)$
is not contained in the moduli space $\mathcal{M}_{0,5}$ (see appendix
B, eq. \ref{eq:moduli space cubical coordinates}). In order to evaluate
functions there, one has to work with a compactification of this space,
obtained by blowing up this point to a line, such that the boundary
transforms from a square to a pentagon in this case. On this space,
one clearly arrives at different points, depending on whether one
approaches the additional line from $L_{1}$ or $L_{2}.$ We refer
to \cite{Bro1} for a detailed, general discussion.

\subsection{Definite integration\label{sub:Period-integrals-on}}

Let us consider cubical coordinates $x_{1},\,...,\, x_{m}$ and convergent
integrals of the form 
\begin{equation}
I=\int_{0}^{1}dx_{m}\frac{q}{\prod_{i}p_{i}^{a_{i}}}f,\label{eq:Period integral on M0,n}
\end{equation}
where $f\in V\left(\Omega_{m}\right),$ $q$ is a polynomial in $x_{m}$,
the $a_{i}\in\mathbb{N}$ and the $p_{i}$ are elements of the set
\begin{equation}
\mathcal{P}_{C}=\left\{ x_{m},\,1-x_{m},\,...,\,1-x_{1}\cdot\cdot\cdot x_{m}\right\} ,\label{eq:P C polynomials}
\end{equation}
 i.e. in the set of denominators of the differential 1-forms in $\Omega_{m}.$
The analytical computation of such integrals with MPL proceeds as
follows:
\begin{itemize}
\item The integral $I$ is expressed as a linear combination of integrals
of the type 
\[
I_{i}=\int_{0}^{1}\omega_{i}f_{i}
\]
 with $\omega_{i}\in\bar{\Omega}_{m}^{F}$, $f_{i}\in V\left(\Omega_{m}\right).$
This is achieved by a combination of partial fraction decompositions
and partial integrations, iteratively lowering the exponents $a_{i}.$
The partial integrations involve the computation of limits of $f$
at $x_{m}=0$ and $x_{m}=1$ as discussed in section \ref{sub:Limits}.
\item For each integral $I_{i}$ the primitive $\tilde{I}_{i}=\int\omega_{i}\otimes f_{i}$
is computed as in section \ref{sub:Primitives}.
\item The limits $\lim_{x_{m}\rightarrow1}\tilde{I}_{i}-\lim_{x_{m}\rightarrow0}\tilde{I}_{i}$
are evaluated. 
\end{itemize}
All of these steps are combined in the procedure \texttt{MPLCubicalIntegrate(f,var,n)}.
The first argument is an integrand as in eq. \ref{sub:Period-integrals-on}
in a set of cubical coordinates $x_{1},\,...,\, x_{m}$ and the second
argument is the last of these variables, $x_{m}.$ The positive integer
$n$ is the number of integrations to be computed. The procedure integrates
from 0 to 1 over the variables $x_{m},\, x_{m-1},\,...,\, x_{m-n+1}$
in this order.

\subsubsection*{Example:}

We consider two families of integrals 
\begin{eqnarray*}
I_{1}(N) & = & \int_{0}^{1}\int_{0}^{1}dx_{1}dx_{2}f_{1}(N)\;\textrm{and}\; I_{2}(N)=\int_{0}^{1}\int_{0}^{1}\int_{0}^{1}dx_{1}dx_{2}dx_{3}f_{2}(N),
\end{eqnarray*}
with integrands 
\begin{eqnarray*}
f_{1}(N) & = & (-1)^{N}\frac{x_{1}^{N}(1-x_{1})^{N}x_{2}^{N}(1-x_{2})^{N}}{(1-x_{1}x_{2})^{N+1}},\\
f_{2}(N) & = & \frac{x_{1}^{N}(1-x_{1})^{N}x_{2}^{2N+1}(1-x_{2})^{N}x_{3}^{N}(1-x_{3})^{N}}{(1-x_{1}x_{2})^{N+1}(1-x_{2}x_{3})^{N+1}}
\end{eqnarray*}
for $N\in\mathbb{N}\cup\{0\}.$

These integrals arise, after a change of variables, from Beukers'
proofs \cite{Beu} of Ap\'ery's theorems \cite{Ape} on the irrationality
of $\zeta(2)$ and $\zeta(3).$ The general role of period integrals
on $\mathcal{M}_{0,n}$ in irrationality proofs for zeta values is
worked out in \cite{Bro4}. MPL can be used to confirm the results
\begin{eqnarray}
I_{1}(N) & = & a_{1}(N)\zeta(2)-b_{1}(N),\nonumber \\
I_{2}(N) & = & 2a_{2}(N)\zeta(3)-2b_{2}(N),\label{eq:Beukers integrals}
\end{eqnarray}
where the sequences of numbers $a_{1}(N),\, b_{1}(N)$ satisfy the
recurrence relation
\[
u_{1}(N)=N^{-2}\left(\left(11N^{2}-11N+3\right)u_{1}(N-1)+\left(N-1\right)^{2}u_{1}(N-2)\right)
\]
with initial conditions $a_{1}(0)=1,\, a_{1}(1)=3,\, b_{1}(0)=0,\, b_{1}(1)=5$
and the numbers $a_{2}(N),\, b_{2}(N)$ satisfy
\[
u_{2}(N)=N^{-3}\left(\left(34N^{3}-51N^{2}+27N-5\right)u_{2}(N-1)-(N-1)^{3}u_{2}(N-2)\right)
\]
with $a_{2}(0)=1,\, a_{2}(1)=5,\, b_{2}(0)=0,\, b_{2}(1)=6.$

For example, setting the first argument \texttt{f} of the procedure
\texttt{MPLCubicalIntegrate} equal to $f_{2}(4),$ we compute%
\footnote{Cubical coordinates are declared internally before each integration,
therefore it is not necessary to use \texttt{MPLCoordinates} here. %
} 
\begin{quote}
\texttt{MPLCubicalIntegrate(f,x{[}3{]},3);}~\\
$-11424695/144+66002\zeta(3)$
\end{quote}
in agreement with eq. \ref{eq:Beukers integrals}. Here the program
integrates over $x_{3},\, x_{2},\, x_{1}.$ Intermediate results can
be obtained by setting the third argument equal to 1 or 2.

\section{Integration over Feynman parameters\label{sec:Integration-over-Feynman}}

In this section, we introduce procedures to express a very general
class of integrals in terms of integrals of the form of eq. \ref{eq:Period integral on M0,n},
such that the procedures of section \ref{sec:Computing-with-iterated}
can be used for their computation. For the sake of concreteness, we
assume an application to Feynman integrals of perturbative quantum
field theory, but the procedures may apply to integrals of a different
context as well. 

We consider integrals of the form 
\begin{equation}
I_{F}=\int_{0}^{\infty}d\alpha_{j}\frac{\prod_{Q_{i}\in\mathcal{Q}}Q_{i}^{\delta_{i}}L_{w}\left(\alpha_{j}\right)}{\prod_{P_{i}\in\mathcal{P}}P_{i}^{\beta_{i}}}\label{eq:Integrals Feynman}
\end{equation}
where $\mathcal{Q},\,\mathcal{P}\subset\mathbb{Q}\left[\alpha_{1},\,...,\,\alpha_{j},\,...,\,\alpha_{N}\right]$
are sets of irreducible polynomials, all $\delta_{i},\,\beta_{i}\in\mathbb{N}\cup\left\{ 0\right\} $
and $L_{w}\left(\alpha_{N}\right)$ is a hyperlogarithm, given by
a word $w$ in differential 1-forms in 
\begin{equation}
\Omega_{F}=\left\{ \frac{d\alpha_{j}}{\alpha_{j}},\,\frac{d\alpha_{j}}{\alpha_{j}-\rho_{i}}\textrm{ where }\rho_{i}=-\frac{P_{i}|_{\alpha_{j}=0}}{\frac{\partial P_{i}}{\partial\alpha_{j}}},\, P_{i}\in\mathcal{P}\right\} .\label{eq:1-forms Feynman}
\end{equation}

Let us compare the integrals of eq. \ref{eq:Integrals Feynman} to
the ones of eq. \ref{eq:Period integral on M0,n}. The essential difference
is between the sets $\mathcal{P}_{C}$ and $\mathcal{P},$ i.e. the
polynomials in the denominator of the integrand and defining the denominators
of the differential 1-forms of the iterated integrals in the numerator
of the integrand. While $\mathcal{P}_{C}$ is a very specific set
of polynomials (eq. \ref{eq:P C polynomials}), we will allow the
set $\mathcal{P}$ to contain polynomials of a much more general type. 

In the following, we consider successive integrations of the type
of eq. \ref{eq:Integrals Feynman} over several variables $\alpha_{\sigma(1)},\,\alpha_{\sigma(2)},\,...,\,\alpha_{\sigma(N)}$
with respect to some permutation $\sigma$ of $\{1,\,...,\, N\}.$
Let $\mathcal{P}^{(1,\,...,\, i)}$ denote the set of polynomials
which plays the role of $\mathcal{P}$ in eq. \ref{eq:Integrals Feynman}
and \ref{eq:1-forms Feynman} for integrands after the first $i$
integrations. In the following, let $\mathcal{P}$ denote the first
such set, before any integrations. Let us call a polynomial \emph{at
most linear} in some variable, if its degree in this variable is either
0 or 1.

Integrals of this type can be computed with MPL, if the following
conditions are satisfied: 
\begin{enumerate}
\item The integral has to be \emph{finite}. Of course, this condition is
failed by many Feynman integrals. However, there are powerful methods
to express divergent Feynman integrals in terms of finite integrals,
e.g. \cite{Pan3,ManPanSch,BroKre}. In the following we assume that
the integral under consideration is either a finite Feynman integral
or one of the finite integrals arising from such a procedure.
\item The integral has to be \emph{linearly reducible} \cite{Bro5}, i.e.
there is an order of integrations $\sigma$ such that every polynomial
in $\mathcal{P}^{(\sigma(1),\,...,\,\sigma(i))}$ is at most linear
in $\alpha_{\sigma(i+1)}$ for all $i=0,\,...,\, N-1.$ Linear reducibility
can be checked before the integration procedure by the algorithms
of \cite{Bro5,Bro6} which are implemented in MPL as discussed in
section \ref{sub:Linear-reduction}.
\item Our implementation of the computation of limits (section \ref{sub:Limits})
implies, that the integral has to be \emph{unramified} (as defined
in \cite{Bro5}) and $\mathcal{P}$ has to satisfy a condition, which
we will call \emph{properly ordered}. In section \ref{sub:Changes-of-variables},
we recall the aspects of our discussion of \cite{BogBro2}, which
make these conditions precise.
\end{enumerate}
Procedures to check these conditions and to compute an integral which
satisfies the conditions are introduced below.

\subsection{Polynomial reduction\label{sub:Linear-reduction}}

For each permutation $\sigma$ on $\{1,\,...,\, N\},$ Brown's polynomial
reduction algorithms \cite{Bro5,Bro6} construct sequences 
\[
S^{\{\sigma(1)\}},\, S^{\{\sigma(1),\,\sigma(2)\}},\,...,\, S^{\{\sigma(1),\,\sigma(2),\,...,\,\sigma(i)\}},
\]
 with $i\leq N,$ of sets of irreducible polynomials, such that 
\begin{equation}
\mathcal{P}^{\left(\sigma(1),\,...,\,\sigma(i)\right)}\subseteq S^{\{\sigma(1),\,...,\,\sigma(i)\}},\, i\in\{1,\,...,\, N\}.\label{eq:upper bound}
\end{equation}
If $S^{\{\sigma(1),\,...,\,\sigma(i)\}}$ contains a polynomial of
degree greater than 1 in $\alpha_{\sigma(i+1)}$ then the set $S^{\{\sigma(1),\,...,\,\sigma(i),\,\sigma(i+1)\}}$
with respect to the order given by $\sigma$ is not constructed. The
initial set $\mathcal{P}$ is called linearly reducible, if there
is a $\sigma$ such that the algorithm succeeds to construct a full
sequence $S^{\{\sigma(1)\}},\,...,\, S^{\{\sigma(1),\,...,\,\sigma(N)\}}.$

We briefly recall the construction of the sets $S^{\{\sigma(1),\,...,\,\sigma(i)\}}.$
Let $S\subset\mathbb{Q}\left[\alpha_{1},\,...,\,\alpha_{N}\right]$
and let $S_{\textrm{irred.}}$ denote the set of all irreducible factors
of all polynomials in $S,$ disregarding constants. Adapting the notation
of \cite{Pan1} we define 
\[
\left[P,\,0\right]_{i}=P|_{\alpha_{i}=0},\,\left[P,\,\infty\right]_{i}=\begin{cases}
\frac{\partial P}{\partial\alpha_{i}} & \textrm{ if }\frac{\partial P}{\partial\alpha_{i}}\neq0,\\
P|_{\alpha_{i}=0} & \textrm{ otherwise,}
\end{cases}\,\textrm{and }\left[P_{j},\, P_{k}\right]_{i}=\frac{\partial P_{j}}{\partial\alpha_{i}}P_{k}|_{\alpha_{i}=0}-\frac{\partial P_{k}}{\partial\alpha_{i}}P_{j}|_{\alpha_{i}=0}.
\]
For sets $S$ all of whose polynomials are at most linear in $\alpha_{i}$,
we define the \emph{simple reduction} \cite{Bro5} of $S$ with respect
to $\alpha_{i}$ by 
\[
S_{i}=\left\{ \left[P,\,0\right]_{i},\,\left[P,\,\infty\right]_{i}:\, P\in S\right\} _{\textrm{irred.}}\cup\left\{ \left[P_{j},\, P_{k}\right]_{i}:\, P_{j},\, P_{k}\in S\right\} _{\textrm{irred.}}.
\]
One defines the \emph{Fubini reduction} \cite{Bro5} of $\mathcal{P}$
with respect to $\sigma$ by
\begin{eqnarray}
S^{\emptyset} & = & \mathcal{P},\nonumber \\
S^{\left\{ \sigma(1),\,...,\,\sigma(i)\right\} } & = & \cap_{k\in\left\{ \sigma(1),\,...,\,\sigma(i)\right\} }S_{k}^{\left\{ \sigma(1),\,...,\,\sigma(i)\right\} \backslash\{k\}},\label{eq:Fubini reduction}
\end{eqnarray}
where on the right-hand side only the sets $S^{\left\{ \sigma(1),\,...,\,\sigma(i)\right\} \backslash\{k\}}$
are considered, which are defined by simple reduction and whose polynomials
are at most linear $\alpha_{k}$. Disregarding monomials and constants,
these sets satisfy eq. \ref{eq:upper bound}. Moreover, for every
permutation $\lambda$ on $\left\{ \sigma(1),\,...,\,\sigma(i)\right\} $
we have 
\[
\mathcal{P}^{\left(\lambda\left(\sigma(1)\right),\,...,\,\lambda\left(\sigma(i)\right)\right)}\subseteq S^{\{\sigma(1),\,...,\,\sigma(i)\}},\, i\in\{1,\,...,\, N\},
\]
assuming that both sides of this relation exist. 

A more refined upper bound was defined in \cite{Bro6} by introducing
\emph{compatibility graphs}. Here we adapt the version defined in
\cite{Pan1}. The above simple reduction is extended to a pair $(S,\, C)$
where $S$ is a set of irreducible polynomials and $C$ is the set
of edges of a graph, whose vertices are the polynomials in $S$. A
pair of two distinct polynomials in $S$ is called \emph{compatible,}
if there is an edge between the corresponding two vertices in $C$.
For a set $S$ with all polynomials at most linear in $\alpha_{i}$
one defines 
\begin{eqnarray*}
\bar{S}_{i} & = & \left\{ \left[P,\,0\right]_{i},\,\left[P,\,\infty\right]_{i}:\, P\in S\right\} _{\textrm{irred.}}\cup\left\{ \left[P_{j},\, P_{k}\right]_{i}:\,\textrm{ compatible pairs }P_{j},\, P_{k}\in S\right\} _{\textrm{irred.}}.
\end{eqnarray*}
In the corresponding compatibility graph $C_{i}$ the vertices are
given by $\bar{S}_{i}$ and the edges are between all pairs of distinct
irreducible factors of $\left[P_{j},\, P_{k}\right]_{i}\cdot\left[P_{k},\, P_{l}\right]_{i}\cdot\left[P_{l},\, P_{j}\right]_{i}$
for every mutually compatible $P_{j},\, P_{k},\, P_{l}\in S\cup\left\{ 0,\,\infty\right\} .$
The auxiliary terms 0 and $\infty$ are each considered compatible
with every element of $S\cup\left\{ 0,\,\infty\right\} .$ The initial
pair of this reduction is $\left(\mathcal{P},\, K_{\mathcal{P}}\right),$
where $K_{\mathcal{P}}$ is the complete graph whose vertices are
the polynomials in $\mathcal{P}.$ We have $\bar{S}_{i}\subseteq S_{i}$
by construction. It was proven in \cite{Pan1}, that the sequence
obtained by this extended simple reduction is an upper bound for the
sets $\mathcal{P}^{\left(\sigma(1),\,...,\,\sigma(i)\right)}$ as
in eq. \ref{eq:upper bound}.

In analogy to the Fubini reduction, we furthermore define
\begin{eqnarray}
\left(S^{\emptyset},\, C^{\emptyset}\right) & = & \left(\mathcal{P},\, K_{\mathcal{P}}\right),\nonumber \\
\left(\bar{S}^{\left\{ \sigma(1),...,\sigma(i)\right\} },\, C^{\left\{ \sigma(1),...,\sigma(i)\right\} }\right) & = & \left(\cap_{k\in\left\{ \sigma(1),...,\sigma(i)\right\} }\bar{S}_{k}^{\left\{ \sigma(1),...,\sigma(i)\right\} \backslash\{k\}},\right.\nonumber \\
 &  & \left.\,\cap_{k\in\left\{ \sigma(1),...,\sigma(i)\right\} }C_{k}^{\left\{ \sigma(1),...,\sigma(i)\right\} \backslash\{k\}}\right)\label{eq:Compatibility graph reduction}
\end{eqnarray}
$ $where the intersection $G_{1}\cap G_{2}$ of graphs $G_{1},\, G_{2}$
is a graph consisting only of the vertices and the edges which belong
to both of the graphs. Note that this construction includes the above
Fubini reduction, if we replace every graph in eq. \ref{eq:Compatibility graph reduction}
with the corresponding complete graph. Therefore we have 
\[
\bar{S}^{\left\{ \sigma(1),\,...,\,\sigma(i)\right\} }\subseteq S^{\left\{ \sigma(1),\,...,\,\sigma(i)\right\} }
\]
by construction. An explicit proof that the sets $\bar{S}^{\left\{ \sigma(1),\,...,\,\sigma(i)\right\} }$
satisfy eq. \ref{eq:upper bound} is missing at this point. However,
these sets have served as correct upper bounds in all known cases
so far (see the remark in sec. 3.6.6 of \cite{Pan1}). 

The program MPL provides the procedure 
\begin{quote}
\texttt{>MPLPolynomialReduction(S,L);}
\end{quote}
where the first argument is the list of polynomials to be reduced
and the second argument is the list of all variables with respect
to which the reduction shall be computed. In a typical Feynman integral
computation, $S$ would contain the Symanzik polynomials and $L$
would be the list of Feynman parameters (cf. the example in appendix
A). The procedure returns a list where each entry corresponds to a
pair $\left(\bar{S}^{\left\{ \sigma(1),\,...,\,\sigma(i)\right\} },\, C^{\left\{ \sigma(1),\,...,\,\sigma(i)\right\} }\right)$
of the above reduction. For example, a typical entry of this list
would be 
\[
\left[\left[\alpha_{1},\,\alpha_{4}\right],\left[\alpha_{2}\alpha_{3}+1,\,\alpha_{2}-\alpha_{3},\,\alpha_{2}\alpha_{5}+\alpha_{3}\right],\left[\left\{ 1,2\right\} ,\left\{ 1,3\right\} \right]\right].
\]
Every entry contains three lists. The first list contains the reduced
variables, the second list contains the polynomials in the remaining
variables (disregarding monomials and constants) and the third list
contains the edges of the compatibility graph, in terms of pairs of
the numbers of compatible polynomials with respect to the given list.
In this sense, the above example stands for the pair $\left(\bar{S}^{\left\{ 1,\,4\right\} },\, C^{\left\{ 1,\,4\right\} }\right)$
with $\bar{S}^{\left\{ 1,\,4\right\} }=\left\{ \alpha_{2}\alpha_{3}+1,\,\alpha_{2}-\alpha_{3},\,\alpha_{2}\alpha_{5}+\alpha_{3}\right\} $
and where the first polynomial of this list is compatible with the
second one and the third one. 

A global, boolean variable \texttt{COMPATIBILITY\_GRAPH} is set true
by default. If the user sets this variable to false, the procedure
\texttt{MPLPolynomialReduction} returns a list corresponding to the
sets $S^{\{\sigma(1),\,...,\,\sigma(i)\}}$ of the Fubini reduction
(eq. \ref{eq:Fubini reduction}).

In both cases, the polynomial reduction algorithm considers all permutations
on $\{1,\,...,\, N\}.$ If the reduction (by at least one of both
algorithms) contains a complete sequence 
\[
S^{\{\sigma(1)\}},\,...,\, S^{\{\sigma(1),\,...,\,\sigma(N)\}}
\]
 (or with $\bar{S}$ instead of $S$) for one such permutation $\sigma,$
linear reducibility is satisfied. 

If this is not the case, the algorithm has stopped the construction
of sets at certain stages, because of polynomials which are non-linear
in some $\alpha_{i}.$ There are several possible scenarios. Firstly,
these polynomials might be spurious, i.e. they belong to the upper
bounds generated by the algorithm, but might not appear in the actual
integrands. In this (rather unlikely) case, the integral can still
be computed. Secondly, the problematic polynomials may belong to $\mathcal{P}^{\left(\sigma(1),\,...,\,\sigma(i)\right)}$,
but it may be possible to restore linearity in some variable by an
appropriate change of variables. Such examples%
\footnote{An example for this case is the massless graph $K_{4}$ with four
on-shell legs, which was found to be irreducible with respect to Feynman
parameters in \cite{BogLue} but computable in terms of harmonic polylogarithms
in \cite{HennSmi}. A change of variables restoring linear reducibility
was found in \cite{Pan3}.%
} are discussed in \cite{Pan3,Pan1}. These changes of variables are
not implemented in MPL so far. Thirdly, it may be the case that such
changes of variables do not exist and that neither iterated integrals
on $\mathcal{M}_{0,n}$ nor hyperlogarithms are sufficient to express
the given integral%
\footnote{In such cases, the use of elliptic polylogarithms as in the computations
\cite{BloVan,BloKerVan,AdaBogWei1,AdaBogWei2} may be appropriate.%
}.

\subsection{Changes of variables \label{sub:Changes-of-variables}}

Let us assume an integrand of the type of eq. \ref{eq:Integrals Feynman}
whose polynomials and hyperlogarithms depend on $N$ Feynman parameters
$\alpha_{1},\,...,\,\alpha_{N}$ and on $M$ kinematical invariants
$s_{1},\,...,\, s_{M}$ (possibly including squared particle masses).
Furthermore, let these kinematical invariants be defined such that
in the momentum region where we want to compute the integral, all
$s_{i}$ take only non-negative, real values. The most convenient
way to compute with such variables in MPL is to treat them like additional
Feynman parameters, which are not integrated out. In this sense, let
us define the $N+M$ \emph{generalized Feynman parameters} $a_{i}$
by 
\[
a_{i}=s_{i}\textrm{ and }a_{M+j}=\alpha_{j}\textrm{ for }i=1,\,...,\, M\textrm{ and }j=1,\,...,\, N.
\]

The transformations of integrands from Feynman parameters to cubical
coordinates and back are discussed in detail in section 4 of \cite{BogBro2},
without explicitly mentioning kinematical invariants. However, the
discussion in \cite{BogBro2} is easily extended to include these
dependences by simply replacing the Feynman parameters by the generalized
Feynman parameters. Making this point explicit here, we recall the
remaining conditions on the integrand. 

Let us assume that the set of polynomials $\mathcal{P}=\left\{ P_{1},\,...,\, P_{m}\right\} $
of the integrand is linearly reducible. Let $T=N+M$ and let us number
the (generalized) Feynman parameters such that from left to right
$a_{T},\, a_{T-1},\,...,\, a_{M+1}$ is the desired order of integrations.
Following the algorithm of \cite{BogBro2}, MPL introduces $m$ cubical
coordinates $x_{1},\,...,\, x_{m}$ by a homomorphism $\phi^{\star}:\,\mathbb{Q}\left(x_{1},\,...,\, x_{m}\right)\rightarrow\mathbb{Q}\left(a_{1},\,..,\, a_{T}\right)$
and expresses the integrand in terms of integrals of the type \ref{eq:Period integral on M0,n},
such that all polynomials and hyperlogarithms depending on the next
integration variable $a_{T}=\alpha_{N}$ are expressed in terms of
cubical coordinates. After integrating over $x_{m},$ the result is
expressed in terms of hyperlogarithms in Feynman parameters again.
As explained in detail in \cite{BogBro2}, these functions $\eta$
have to satisfy a certain condition at the tangential basepoint corresponding
to consecutive limits, for which we define the short-hand notation
$\lim_{a\rightarrow0_{T-1}}$ by 
\begin{equation}
\lim_{a\rightarrow0_{T-1}}\eta=\lim_{a_{1}\rightarrow0}...\lim_{a_{T-1}\rightarrow0}\eta.\label{eq:condition tangential basepoint hyperlogs}
\end{equation}
These limits in the space of Feynman parameters correspond to limits
in cubical coordinates, in the sense that 
\[
\lim_{x_{p}\rightarrow c}g=\lim_{a\rightarrow0_{T-1}}\phi^{\star}g
\]
for any rational function $g$ in the $x$-coordinates. Here $\lim_{x_{p}\rightarrow c}g$
is the short-hand notation defined by

\begin{equation}
\lim_{x_{p}\rightarrow c}g=\lim_{x_{p(1)}\rightarrow c_{p(1)}}...\lim_{x_{p(m-1)}\rightarrow c_{p(m-1)}}g\label{eq:multiple limit cubical coordinates}
\end{equation}
where $p$ is a particular permutation on $\{1,\,..,\, m-1\}$ and
$c_{i}=\lim_{a\rightarrow0_{T-1}}x_{i}$ for $i=1,\,...,\, m-1.$
As we take limits with respect to all remaining variables here, the
$c_{i}$ are real numbers. 

In MPL, limits as in eq. \ref{eq:multiple limit cubical coordinates}
can be computed by use of the procedures discussed in section \ref{sub:Limits},
if two conditions are satisfied. Firstly, the limit has to be computed
at a corner-point of the unit-cube

\[
\mathbb{R}_{\textrm{cube}}^{m-1}=\left\{ \left(x_{1},\,...,\, x_{m-1}\right)\in\mathbb{R}^{m-1}|0\leq x_{i}\leq1,\, i=1,\,...,\, m-1\right\} ,
\]
i.e. $c_{i}\in\left\{ 0,\,1\right\} $ for $i=1,\,...,\, m-1.$ Secondly,
the limit has to be computed by approaching this point from inside
this cube. 

We introduce two conditions on the set $\mathcal{P},$ such that these
conditions on the limits are satisfied. For each $P_{i}\in\mathcal{P}$
let us define $\rho_{i}=-\frac{P_{i}|_{\alpha_{N}=0}}{\frac{\partial P_{i}}{\partial\alpha_{N}}}$
and consider the space of generalized Feynman parameters 
\[
\mathbb{R}^{T-1}=\left\{ \left(a_{1},\,...,\, a_{T-1}\right)\in\mathbb{R}^{T-1}|0\leq a_{i},\, i=1,\,...,\, T-1\right\} .
\]

As a first condition, let there be an open region $\Lambda\subset\mathbb{R}^{T-1}$
with all points satisfying $0\leq a_{T-1}\ll a_{N-2}\ll...\ll a_{1}\ll\epsilon,$
where $x\ll y$ denotes $x<y^{K},$ for a sufficiently small number
$\epsilon$ and a sufficiently large number $K$, such that everywhere
in $\Lambda$ we have

\begin{equation}
0>\rho_{m}>\rho_{1}>\rho_{2}>...>\rho_{m-2}>\rho_{m-1}.\label{eq:assumption 1}
\end{equation}
Here we have already numbered the $P_{i}$ according to this unique
order. If such a region $\Lambda$ exists, let us call the integral
\emph{properly ordered}. 

The change of coordinates introduced in \cite{BogBro2} is defined
as 
\[
\phi^{\star}(x_{m})=\frac{a_{T}}{a_{T}-\rho_{m}},\,\phi^{\star}(x_{m-1})=\xi_{m-1}\textrm{ and }\phi^{\star}(x_{k})=\frac{\xi_{k}}{\xi_{k+1}}\textrm{ for }1\leq k\leq m-2,
\]
where $\xi_{i}=1-\frac{\rho_{m}}{\rho_{i}}$ for $i=1,\,...,\, m-1.$
As a consequence of eq. \ref{eq:assumption 1}, we have $0<\xi_{i}<1$
and $\xi_{i}<\xi_{i+1}$ and therefore $0<x_{i}<1,$ $i=1,\,...,\, m-1,$
everywhere in $\Lambda.$ Therefore, if the integral is properly ordered,
the limit eq. \ref{eq:multiple limit cubical coordinates} is approached
from inside $\mathbb{R}_{\textrm{cube}}^{m-1}.$

As a second condition, we assume that the set $\left\{ \rho_{1},\,...,\,\rho_{m}\right\} $
is \emph{unramified} \cite{Bro5}, i.e. 
\[
\lim_{a\rightarrow0_{T-1}}\rho_{i}\in\{0,\,-1,\,\infty\}\textrm{ for }i=1,\,...,\, m.
\]
 In this case, we also call the integral unramified. Together with
the condition to be properly ordered, this implies the desired condition
$c_{i}\in\left\{ 0,\,1\right\} $ for $i=1,\,...,\, m-1.$

Note that it may strongly depend on the choice of kinematical invariants
and the region of the momentum space, whether these conditions are
satisfied. We also note, that while the choice of the order of $a_{T},\,...,\, a_{M+1}$
is at least partially restricted by the condition of linear reducibility,
the order of the kinematical invariants $a_{M},\,...,\, a_{1}$ is
arbitrary. If this freedom in the choice of the orders and in the
choice of kinematical invariants can not be exploited to write the
integrand in a form such that proper ordering and unramifiedness are
satisfied, the use of iterated integrals on $\mathcal{M}_{0,n}$ or
hyperlogarithms may still be possible.

\subsection{The computation of Feynman integrals with MPL}

Let us now turn to the computation of a finite integral in Feynman
parameters with MPL. We choose a letter to denote the generalized
Feynman parameters, say $a_{i}$. In the Maple worksheet, this choice
is declared by the Maple command
\begin{quote}
\texttt{>defform(a=0):}
\end{quote}
From here on, all \texttt{a{[}i{]}} are treated as variables and all
\texttt{d(a{[}i{]})} are treated as differential 1-forms by Maple. 

For a given set $\mathcal{P}$ of polynomials in the integrand of
the type of eq. \ref{eq:Integrals Feynman}, we check with \texttt{MPLPolynomialReduction}
whether the integrand is linearly reducible (see section \ref{sub:Linear-reduction}).
Here let us assume that linear reducibility is satisfied. For convenience,
we number the integration variables such that an allowed order of
integrations is $a_{T},\, a_{T-1},...,\, a_{M+2},\, a_{M+1}.$ Furthermore
we number the $M$ kinematical invariants according to some arbitrary
order. Based on these orders, we define the vector of all generalized
Feynman parameters $a=\left(a_{T},\,,...,\,,\, a_{M+1},\, a_{M},\,...,\, a_{1}\right).$ 

The question remains, whether the integral is furthermore properly
ordered and unramified with respect to the order in $a.$ This can
be checked with the MPL command
\begin{quote}
\texttt{>MPLCheckOrder(reduction, a{[}1..k{]}, a);}
\end{quote}
Here the first argument is the list generated by \texttt{MPLPolynomialReduction},
the second argument is a list of the variables in $a$ to be integrated
out and the last argument is the list \texttt{a={[}a{[}T{]},...,a{[}1{]}{]}}
of all components of $a.$ In both lists, the entries are ordered
according to the chosen vector $a.$ 

If the latter procedure confirms the conditions to be satisfied, we
can use the order of $a$ to compute the integral%
\footnote{If one of the conditions fails, the freedom of choices with respect
to kinematical invariants as mentioned in section \ref{sub:Changes-of-variables}
may be used to try a different $a.$%
}. This is done with the command

\texttt{>MPLFeynmanIntegrate(integrand, a{[}1..k{]}, a);}

Here the first argument is the integrand of the type of eq. \ref{eq:Integrals Feynman}.
Note that the numerator of the integrand has to be a linear combination
of logarithms and hyperlogarithms in bar-notation (using \texttt{bar(...)}),
if such functions occur. The second and third argument are the same
as in the previous procedure. We obtain the final result by integrating
over $k=T-M$ variables and intermediate results by chosing $k<T-M.$
The results are given in terms of hyperlogarithms $L$ in the remaining
generalized Feynman parameters, vanishing in the limit $\lim_{a\rightarrow0_{T}}L$
by definition.

The Feynman parametric version of the integral is sometimes expressed
with a $\delta(H)$ in the integrand, where $H$ is some hypersurface
in the space of the integration variables, to be chosen freely, according
to the famous Cheng-Wu theorem \cite{CheWu}. We always choose $H=1-a_{M+1},$
which implies that we integrate over the first $k=T-M-1$variables
with the above command and take the limit $\lim_{a_{M+1}\rightarrow0}$. 

Detailed examples of Feynman integral computations with MPL are presented
in the manual and in appendix A.

\section{Conclusions\label{sec:Conclusions}}

We have introduced the computer program MPL for computations with
iterated integrals on moduli spaces of curves of genus zero with $n$
ordered marked points with Maple. The program includes procedures
for the computation of the symbol map, the unshuffle map, the total
differentiation, the derivation of primitives and certain limits.
It provides the analytical computation of a very general class of
integrals which can be understood as period integrals on the mentioned
moduli spaces.

By additional procedures, deriving appropriate changes of variables,
the program furthermore supports the computation of Feynman integrals.
It provides the automated computation of finite, linearly reducible
integrals over Feynman parameters if the order of integrations satisfies
unramifiedness and properly ordered polynomials at a tangential basepoint.

\section*{Acknowledgements}

I thank Francis Brown for introducing me to the subject of this article,
for crucial advice at various stages of writing the program and very
useful remarks on the manuscript. I am grateful to Erik Panzer for
his help, particularly on the application to Feynman integrals, and
also for very useful remarks on the manuscript. I am also indebted
to Dirk Kreimer for his strong support and encouragement for the project.
This work was partly supported by Deutsche Forschungsgemeinschaft.
The graphs were drawn using \cite{HahLan}.

\section*{Appendix A: An example of a Feynman integral computation}

\begin{figure}
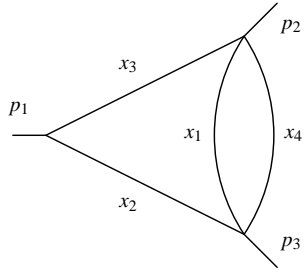

\selectlanguage{german}%
\begin{centering}
{\scriptsize{\begin{feynartspicture}(110,110)(1,1) 

\FADiagram{}
\FAProp(0.,10.)(2.5,10.)(0.,){Straight}{0} \FALabel(0.5,11.52)[b]{$p_1$} \FAProp(2.5,10.)(17.5,17.5)(0.,){Straight}{0} \FALabel(9.48,14.77)[br]{$x_3$} \FAProp(17.5,17.5)(17.5,2.5)(0.3,){Straight}{0} \FALabel(14.43,10.)[r]{$x_1$} \FAProp(17.5,17.5)(17.5,2.5)(-0.3,){Straight}{0} \FALabel(20.57,10.)[l]{$x_4$} \FAProp(17.5,2.5)(2.5,10.)(0.,){Straight}{0} \FALabel(9.48,5.23)[tr]{$x_2$} \FAProp(17.5,17.5)(20.,20.)(0.,){Straight}{0} \FALabel(20.27,18.73)[tl]{$p_2$} \FAProp(17.5,2.5)(20.,0.)(0.,){Straight}{0} \FALabel(20.27,1.27)[bl]{$p_3$} 
\end{feynartspicture}}}
\par\end{centering}{\scriptsize \par}

\selectlanguage{english}%
\caption{A two-loop triangle graph}

\end{figure}

As an example application for the procedures introduced in section
\ref{sec:Integration-over-Feynman}, let us consider the massless,
off-shell Feynman graph of figure 1 whose integral was already computed
in \cite{ChaDuh}. We denote the Feynman parameters by $x_{1},\, x_{2},\, x_{3},\, x_{4}$
and the incoming external momenta by $p_{1},\, p_{2},\, p_{3}.$ The
Symanzik polynomials (see e.g. \cite{BogWei2}) of the graph are 
\begin{eqnarray}
\mathcal{U} & = & x_{1}x_{4}+\left(x_{1}+x_{4}\right)\left(x_{2}+x_{3}\right),\label{eq:U}\\
\mathcal{F} & = & -p_{1}^{2}x_{2}x_{3}\left(x_{1}+x_{4}\right)-p_{2}^{2}x_{1}x_{3}x_{4}-p_{3}^{2}x_{1}x_{2}x_{4}.\nonumber 
\end{eqnarray}
Expressing the kinematical dependences by two variables $x_{5}$ and
$x_{6},$ defined by (cf. \cite{ChaDuh}) 
\[
\frac{p_{2}^{2}}{p_{1}^{2}}=\left(1+x_{5}\right)\left(1+x_{6}\right)\textrm{ and }\frac{p_{3}^{2}}{p_{1}^{2}}=x_{5}x_{6},
\]
we furthermore define a slightly modified second Symanzik polynomial
\begin{eqnarray}
\tilde{\mathcal{F}} & = & -\frac{\mathcal{F}}{p_{1}^{2}}=x_{2}x_{3}\left(x_{1}+x_{4}\right)+x_{1}x_{3}x_{4}\left(1+x_{5}\right)\left(1+x_{6}\right)+x_{1}x_{2}x_{4}x_{5}x_{6}\label{eq:F}
\end{eqnarray}
and consider the Feynman integral around $D=4-2\epsilon$ dimensions
\[
I=\int_{x}\delta\left(H\right)\mathcal{U}^{3\epsilon-2}\tilde{\mathcal{F}}^{-2\epsilon}.
\]
Here we omitted a trivial prefactor $\left(-p_{1}^{2}\right)^{2\epsilon}\Gamma\left(2\epsilon\right)$
and we introduced $\int_{x}$ as a short-hand notation for $\prod_{i=1}^{4}\left(\int_{0}^{\infty}dx_{i}\right).$
In the following, we consider the region in momentum space, where
$x_{5}>0,\, x_{6}>0,$ i.e. the set $\left\{ x_{1},\,...,\, x_{6}\right\} $
are generalized Feynman parameters in the sense of section \ref{sub:Changes-of-variables}.
For the hyperplane $H$ we still have a freedom of a choice to be
made below. 

Let us at first express this divergent integral in terms of finite
ones. Following Panzer's strategy of analytic regularization \cite{Pan3},
we define a differential operator 
\[
D_{\left\{ x_{1},\, x_{4}\right\} }=\frac{1}{\epsilon}\left(\left(\epsilon-2\right)-x_{1}\frac{\partial}{\partial x_{1}}-x_{4}\frac{\partial}{\partial x_{4}}\right)
\]
and obtain $I=\frac{1}{\epsilon}\tilde{I}$ with 
\begin{eqnarray*}
\tilde{I} & = & \epsilon\int_{x}\delta\left(H\right)D_{\left\{ x_{1},\, x_{4}\right\} }\left(\mathcal{U}^{3\epsilon-2}\tilde{\mathcal{F}}^{-2\epsilon}\right)\\
 & = & \int_{x}\delta\left(H\right)\frac{\mathcal{U}^{3\epsilon-2}\tilde{\mathcal{F}}^{-2\epsilon}\left(P+\epsilon Q\right)}{\mathcal{U}\mathcal{F}}.
\end{eqnarray*}
Here $P$ and $Q$ are polynomials in $x_{1},\,...,\, x_{6}$, implicitly
defined by this relation. The integral $\tilde{I}$ is finite at $\epsilon=0$.
By expanding its integrand at this point we obtain 
\begin{eqnarray}
\tilde{I} & = & \tilde{I}^{(0)}+\epsilon\tilde{I}^{(1)}+\epsilon^{2}\tilde{I}^{(2)}+\mathcal{O}\left(\epsilon^{3}\right),\nonumber \\
\tilde{I}^{(0)} & = & \int_{x}\delta\left(H\right)\frac{P}{\mathcal{U}^{3}\tilde{\mathcal{F}}},\nonumber \\
\tilde{I}^{(1)} & = & \int_{x}\delta\left(H\right)\frac{Q+P\left(3\ln\left(\mathcal{U}\right)-2\ln\left(\tilde{\mathcal{F}}\right)\right)}{\mathcal{U}^{3}\tilde{\mathcal{F}}},\nonumber \\
\tilde{I}^{(2)} & = & \int_{x}\delta\left(H\right)\frac{Q\left(6\ln\left(\mathcal{U}\right)-4\ln\left(\tilde{\mathcal{F}}\right)\right)+P\left(9\ln\left(\mathcal{U}\right)^{2}+4\ln\left(\tilde{\mathcal{F}}\right)^{2}-12\ln\left(\mathcal{U}\right)\ln\left(\tilde{\mathcal{F}}\right)\right)}{2\mathcal{U}^{3}\tilde{\mathcal{F}}}.\label{eq:Beispielintegrale}
\end{eqnarray}

In the Maple worksheet, we begin the computation with 
\begin{quote}
\texttt{>defform(x=0):}
\end{quote}
From here on, Maple considers every \texttt{x{[}i{]}} as a variable
and every \texttt{d(x{[}i{]})} as a differential 1-form. 

In order to check, whether the integrals $\tilde{I}^{(0)},\,\tilde{I}^{(1)},\,\tilde{I}^{(2)}$
can be computed with MPL, we begin with a polynomial reduction of
the set $S=\left\{ \mathcal{U},\,\tilde{\mathcal{F}}\right\} $. The
reduction is obtained by
\begin{quote}
\texttt{>R:=MPLPolynomialReduction({[}U,F{]},{[}x{[}1{]},x{[}2{]},x{[}3{]},x{[}4{]}{]});}
\end{quote}
The list in the first argument contains the polynomials to be reduced
and the second argument is the list of integration variables in arbitrary
order. The output \texttt{R} of this command is a list where each
entry corresponds to a $\bar{S}^{\left\{ \sigma(1),\,...,\,\sigma(k)\right\} }$
as explained in section \ref{sub:Linear-reduction}. It is easy to
check, that there are permutations $\sigma$ such that a complete
reduction 
\[
\left\{ \bar{S}^{\left\{ \sigma(1)\right\} },\,\bar{S}^{\left\{ \sigma(1),\,\sigma(2)\right\} },\,\bar{S}^{\left\{ \sigma(1),\,\sigma(2),\,\sigma(3)\right\} },\,\bar{S}^{\left\{ \sigma(1),\,\sigma(2),\,\sigma(3),\,\sigma(4)\right\} }\right\} 
\]
is contained in this list. One of these permutations defines the order
$x_{1},\, x_{4},\, x_{3},\, x_{2}.$ Defining an order on all generalized
Feynman parameters accordingly, we choose $a=\left(x_{1},\, x_{4},\, x_{3},\, x_{2},\, x_{5},\, x_{6}\right)$
and check, whether the integrand is unramified and properly ordered
with respect to this order by
\begin{quote}
\texttt{>a:={[}x{[}1{]},x{[}4{]},x{[}3{]},x{[}2{]},x{[}5{]},x{[}6{]}{]}:}~\\
\texttt{>MPLCheckOrder(R, a{[}1..4{]}, a);}
\end{quote}
As all conditions are satisfied, the integral can be computed with
respect to the chosen order. Now we choose the hyperplane $H$ to
be $1-x_{2}=0.$ 

We have to express all (products of) hyperlogarithms in the integrand
as linear combinations of iterated integrals in bar-notation. For
the given example, it is sufficient to re-write the integrands of
eq. \ref{eq:Beispielintegrale} by use of 
\[
\ln(X)=\left[\frac{d(X)}{X}\right]\textrm{ and }\mbox{\textrm{ln}}(X)\mbox{\textrm{ln}}(Y)=\left[\frac{d(X)}{X}|\frac{d(Y)}{Y}\right]+\left[\frac{d(Y)}{Y}|\frac{d(X)}{X}\right]
\]
where the iterated integrals are expressed by use of \texttt{bar(...)}
in the Maple worksheet (see section \ref{sub:Iterated-integrals}).
With these integrands, we can now compute the first three integrations
with
\begin{quote}
\texttt{>MPLFeynmanIntegrate(integrand, a{[}1..3{]}, a);}
\end{quote}
The output is the result after integration over $x_{1},\, x_{4},\, x_{3}$
and according to our choice of $H$, we finally set $x_{2}=1$ in
this result. Here, all of these results are rational functions of
$x_{2},$ so we can simply use%
\footnote{If one of the results would be an iterated integral depending on $x_{2},$
we would apply the command \texttt{MPLLimit} instead.%
}
\begin{quote}
\texttt{>eval(\%, x{[}2{]}=1);}
\end{quote}
We obtain the results 
\begin{eqnarray*}
\tilde{I}^{(0)} & = & 1,\\
\tilde{I}^{(1)} & = & 5,\\
\tilde{I}^{(2)} & = & \frac{1}{x_{6}-x_{5}}\left(2x_{5}(1+x_{6})\left(\left[\frac{d\left(x_{6}\right)}{1+x_{6}}\right]\left[\frac{d\left(x_{5}\right)}{x_{5}}\right]+\left[\frac{d\left(x_{5}\right)}{x_{5}}|\frac{d\left(x_{5}\right)}{1+x_{5}}\right]\left[\frac{d\left(x_{5}\right)}{1+x_{5}}|\frac{d\left(x_{5}\right)}{x_{5}}\right]\right)\right.\\
 &  & \left.-2x_{6}(1+x_{5})\left(\left[\frac{d\left(x_{5}\right)}{1+x_{5}}\right]\left[\frac{d\left(x_{6}\right)}{x_{6}}\right]+\left[\frac{d\left(x_{6}\right)}{x_{6}}|\frac{d\left(x_{6}\right)}{1+x_{6}}\right]+\left[\frac{d\left(x_{6}\right)}{1+x_{6}}|\frac{d\left(x_{6}\right)}{x_{6}}\right]\right)\right)\\
 &  & -3\zeta(2)+19.
\end{eqnarray*}

Here all bar-terms can be interpreted as logarithms and dilogarithms
in the kinematical invariants $x_{5},\, x_{6}.$ 

The same integrals are as well used as an example in the tutorial-worksheet
of Panzer's program \texttt{HyperInt} \cite{Pan2} and the reader
may find it instructive to compare the expressions. MPL provides the
command \texttt{MPLHlogToBar} to automatically express the function
\texttt{Hlog} of \texttt{HyperInt} in terms of our bar-notation. 

As a final remark, we briefly recall, why MPL in general returns the
result of integrations over Feynman parameters in terms of hyperlogarithms.
Internally, the program introduces cubical coordinates for each integration
variable and computes the integrals with the procedures introduced
in section \ref{sec:Computing-with-iterated}. However, as we are
interested to see the result of the integrations expressed in terms
of the kinematical invariants and possibly remaining Feynman parameters
(such as $x_{2}$ here), which we usually cannot interpret as cubical
coordinates of some $\mathcal{M}_{0,n},$ the program has to return
the result in terms of hyperlogarithms in these parameters. If the
integrand in Feynman parameters and kinematical invariants would already
be of the form of eq. \ref{eq:Period integral on M0,n} from the beginning,
we could simply declare these parameters as cubical coordinates and
compute the integrations by use of \texttt{MPLCubicalIntegrate} as
in section \ref{sub:Period-integrals-on}. Usually, due to the complexity
of Symanzik polynomials, this is not the case and a change of variables
is needed.

\section*{Appendix B: A quick introduction to the spaces $\mathcal{M}_{0,n}$}

In the literature on Feynman integrals, moduli spaces of curves are
not a very common topic. In this article we have frequently mentioned
``moduli spaces $\mathcal{M}_{0,n}$ of curves of genus zero with
$n$ ordered, marked points'', without explaining what this phrase
means. Without assuming familiarity with moduli spaces, we have introduced
the corresponding class of iterated integrals by an ad-hoc definition
in section \ref{sec:Computing-with-iterated}. However, it may be
more satisfying to see, how the differential 1-forms in cubical coordinates
arise and what the underlying spaces look like. Here we try to give
a very basic introduction.

Speaking very generally, a moduli space is a device for the classification
of certain objects with respect to equivalence relations. For a given
set of objects, a moduli space is constructed, such that each point
of the space corresponds to an equivalence class in the set. (See
\cite{HarMor} for a very general discussion of such moduli problems.)
In the case of moduli spaces of curves, the objects are algebraic
curves and each point of the moduli space corresponds to an isomorphism
class of such curves. 

In order to keep technicalities at a minimum here, we make use of
the fact, that there is a canonical bijection between the isomorphism
classes of algebraic curves and the isomorphism classes of Riemann
surfaces (see e.g. \cite{Mum}). As a consequence, the moduli spaces
of algebraic curves and of Riemann surfaces are the same, and we restrict
ourselves to the terminology of Riemann surfaces here. We recall that
a Riemann surface is a connected, complex analytic manifold of (complex)
dimension one and its genus is, pictorially speaking, the number of
its 'handles', i.e. 0 for a sphere, 1 for a torus, 2 for a frame of
a pair of glasses and so on.

It is instructive to consider the case of genus 1 at first. For every
Riemann surface $S$ of genus 1 there is a complex number $j(S)$
(the $j$-invariant of elliptic curves) such that two Riemann surfaces
$S_{1},\, S_{2}$ are isomorphic, if and only if $j\left(S_{1}\right)=j\left(S_{2}\right).$
In other words, for each isomorphism class of Riemann surfaces of
genus 1, there is a unique complex number. Moreover it is known that
for any $\lambda\in\mathbb{C}$, there is an isomorphism class $\mathcal{C}$
of such curves, such that $j\left(S\right)=\lambda$ for $S\in\mathcal{C}.$
Therefore the moduli space $\mathcal{M}_{1}$ of curves (or Riemann
surfaces) of genus 1 is simply the complex plane $\mathbb{C}$, parametrized
by $j.$ Usually this moduli space is viewed as the affine line $\mathbb{A}_{j}^{1}.$

The case of genus 0 is even more simple. Every Riemann surface of
genus 0 is isomorphic to the Riemann sphere $\mathbb{C}\cup\left\{ \infty\right\} .$
Therefore there is only one isomorphism class and the moduli space
$\mathcal{M}_{0}$ of curves (or Riemann surfaces) of genus 0 is only
one point.

At low genus, more interesting moduli spaces can be constructed by
adding data to the Riemann surfaces. One way to do so is by marking
some of their points. In the case of genus 0, let us consider $n$
distinct, ordered points $z_{1},\, z_{2},\,...,\, z_{n}$ on a Riemann
sphere $S.$ By saying that these points are marked, we say $S$ with
these marked points is only isomorphic to a Riemann sphere $S'$ with
ordered, marked points $z_{1}^{\prime},\, z_{2}^{\prime},\,...,\, z_{n}^{\prime},$
if an isomorphism between these Riemann shperes exists, which maps
$z_{i}$ to $z_{i}^{\prime}$ for $i=1,\,...,\, n.$ Due to this restriction,
there will clearly be such spheres which are not isomorphic to each
other and we obtain non-trivial moduli spaces.

We recall that an automorphism is an isomorphism of a manifold onto
itself. For the construction of moduli spaces of curves of genus 0,
the isomorphisms defining the classes corresponding to the points
of these spaces are the automorphisms of the Riemann sphere. These
are the M\"obius transformations 
\[
z\mapsto\frac{\alpha z+\beta}{\gamma z+\delta}\textrm{ with }\alpha,\,\beta,\,\gamma,\,\delta\in\mathbb{C},\,\alpha\delta-\beta\gamma\neq0,
\]
which form the M\"obius group denoted $\textrm{PGL}_{2}\left(\mathbb{C}\right).$ 

Now, on a formal level, we have all ingredients to define the moduli
space $\mathcal{M}_{0,n}$ of curves of genus 0 with $n$ ordered,
marked points: 
\[
\mathcal{M}_{0,n}\left(\mathbb{C}\right)=\left\{ \left(z_{1},\,...,\, z_{n}\right)\in\mathbb{C}^{n}\cup\left\{ \infty\right\} \textrm{ distinct}\right\} /\textrm{PGL}_{2}\left(\mathbb{C}\right).
\]
Here it is constructed as a space of $n$ complex coordinates, given
by the marked points on the Riemann sphere $\mathbb{C}\cup\left\{ \infty\right\} ,$
modulo the automorphisms. Due to the condition that all points are
distinct, $z_{i}\neq z_{j}$ for $i,\, j=1,\,...,\, n,$ all diagonals
are excluded from this space.

In this definition, due to the division by the automorphisms $\textrm{PGL}_{2}\left(\mathbb{C}\right),$
it may still not be obvious, what this space really looks like. In
fact, there is only one property of $\textrm{PGL}_{2}\left(\mathbb{C}\right)$
which we have to understand: For any Riemann sphere with $n$ ordered,
marked points, there is an isomorphism in $\textrm{PGL}_{2}\left(\mathbb{C}\right)$
which maps three of the marked points to the points 0, 1 and $\infty.$
As a consequence, if we choose $n\leq3,$ again all spheres are isomorphic
and we obtain a trivial moduli space with only one point. 

The first non-trivial space is $\mathcal{M}_{0,4}.$ We can map three
of the marked points to 0, 1, $\infty$ by an element of $\textrm{PGL}_{2}\left(\mathbb{C}\right).$
The fourth point is allowed to take every complex value except these
three, and defines a new isomorphism class for each new value. In
this sense, the fourth point parametrizes the moduli space, which
therefore has one complex dimension. Introducing one further marked
point we obtain $\mathcal{M}_{0,5}$ as a space parametrized by two
complex variables, excluding the points where they are equal to 0,
1, $\infty$ or equal to each other. 

Generalizing this concept to $n$ marked points, we use the mentioned
property of $\textrm{PGL}_{2}\left(\mathbb{C}\right)$ to always set
$z_{1}=0,\, z_{n-1}=1,\, z_{n}=\infty$ and we introduce the so-called
simplicial coordinates 
\[
t_{1}=z_{2},\, t_{2}=z_{3},\,...,\, t_{n-3}=z_{n-2}.
\]
In terms of these new coordinates, we obtain 
\[
\mathcal{M}_{0,n}\cong\left\{ \left(t_{1},\,...,\, t_{n-3}\right)\in\mathbb{C}^{n-3}|t_{i}\neq t_{j}\textrm{ and }t_{i}\neq\left\{ 0,\,1\right\} \textrm{ for }i,\, j=1,\,...,\, n-3\right\} .
\]
This construction is already suitable for practical computations.
The division by $\textrm{PGL}_{2}\left(\mathbb{C}\right)$ has been
made explicit and we see, that the space has $n-3$ dimensions. By
a further change of variables 
\[
x_{1}=\frac{t_{1}}{t_{2}},\, x_{2}=\frac{t_{2}}{t_{3}},\,...,\, x_{n-4}=\frac{t_{n-4}}{t_{n-3}},\, x_{n-3}=t_{n-3}
\]
we finally arrive at 
\begin{equation}
\mathcal{M}_{0,n}\cong\left\{ \left(x_{1},\,...,\, x_{n-3}\right)\in\mathbb{C}^{n-3}|x_{i}x_{i+1}...x_{j}\neq\left\{ 0,\,1\right\} \textrm{ for all }1\leq i\leq j\leq n-3\right\} .\label{eq:moduli space cubical coordinates}
\end{equation}
The variables $x_{i}$ are the cubical coordinates used throughout
this article. Due to the above condition to the products $x_{i}x_{i+1}...x_{j}$
we have well-defined differential 1-forms 
\[
\frac{dx_{j}}{x_{j}}\textrm{ and }\frac{d\left(x_{i}x_{i+1}...x_{j}\right)}{x_{i}x_{i+1}...x_{j}-1}\textrm{ for }1\leq i\leq j\leq n-3
\]
on this space.

As a final remark, let us mention that in section \ref{sub:Limits}
we have discussed limits at points on the hypersurfaces, which are
excluded from $\mathcal{M}_{0,n}$ by the above definition. In this
context, we have to work with a compactification $\mathcal{\bar{M}}_{0,n}$
of the space. For this and other advanced issues on $\mathcal{M}_{0,n}$,
we refer to \cite{Bro1} and references therein.

\end{document}